\documentclass [12pt,fleqn]{article}
\usepackage{graphicx}
\def \SC {superconductivity\ }
\def \SCg {superconducting\ }
\def \AF {antiferromagnetism\ }
\def \AFc {antiferromagnetic\ }
\begin{document}
\title{\bf A dynamical mean-field theory approach to
superconductivity and\\ antiferromagnetism in a strongly correlated
electron\\ system.}

\author{H. Watanabe$^1$ and S. Doniach$^{1,2}$\\
{\em Departments of Applied Physics$^1$ and Physics$^2$}\\
{\em Stanford University, Stanford, CA 94305, U.S.A.}}
\maketitle
\begin{abstract}
\baselineskip .25in \indent In this paper, we present the results of 
numerical studies of superconductivity and antiferromagnetism in a strongly correlated electron
system. To do this we construct a Hubbard model on a lattice of self-consistently embedded multi-site
clusters (in practice two sites) by means of a dynamical mean-field theory in which intra-cluster
dynamics is treated essentially exactly. We show that a class of characteristic features which have
been seen in the excitation spectra of high-$T_{c}$ cuprates including the pseudogap and the spin-flip
resonance feature seen in neutron scattering studies, as well as their interplay with the onset of a
pairing correlations, can be captured within a dynamical mean-field theory in which short-wavelength
dynamics are rigorously treated. Thus we infer that the observation of the neutron scattering resonance
in the \SCg state of the cuprate superconductors does not appear to be directly tied to their
quasi-2D character.
\\ 
\indent Although our approach is defined strictly in terms of fermion degrees of freedom, 
we show that we can readily identify the emergence of effective low energy bosonic degrees of freedom
in the presence of a well-defined broken symmetry phase as long as their dynamics are dominated by
short-range, short-wavelength fluctuations. Our exact calculations reveal that the dynamics of the
spin degrees of freedom and the onset of \SC are strongly entangled.  In particular, the dynamics of
staggered spin degrees of freedom builds up coherence and a resonance-like sharp feature as \SCg
pairing correlations set in (this feature diminishes in the normal phase). At the same time a spin gap
develops in the staggered spin susceptibility. Under conditions
of superconducting broken symmetry our approach thus extends static BCS mean field theory to provide
an exact treatment of quantum fluctuations of the BCS order parameter within self-consistent dynamical
mean-field theory.  We find that both equilibrium and dynamical properties of our model can provide a
consistent interpretation of experimental observations.\\ 

\end{abstract} 

\baselineskip 18pt {\bf I. INTRODUCTION} \\\\ \indent Understanding the physics 
of strongly correlated electron systems has been one of the most difficult
challenges in condensed matter physics for several decades. The importance
of electron-electron correlations in high-Tc superconductors was
recognized \cite{anderson1} soon after its discovery and much progress
has been made in understanding its phenomenology. As a result of 
intensive efforts in both theoretical and experimental studies, the detailed 
nature of the microscopic mechanism for high-Tc is starting to become clearer 
and the idea that the \SCg instability is driven by Coulomb interactions is 
becoming more widely accepted, although it still remains controversial.\\ 

Besides undergoing a \SCg transition at high temperatures, the cuprate 
superconductors exhibit other features characteristic of strongly correlated 
systems including a pseudogap, spin fluctuation resonance, and non-Fermi liquid 
behavior in the normal state, some of which are also found in other transition metal 
oxides.\\ 

\indent The key motivation of the present paper is to use an extended local 
dynamical mean field approach to study the effects of strong correlations on the 
\SCg instability. By going beyond the static mean field BCS treatment we are 
able to show that coulomb correlations can stabilize the \SCg state and also 
lead to a relationship between the onset of \SC (or more generally, pairing 
fluctuations) and the spin dynamics similar to that seen experimentally. We are also able to study 
the onset of the pseudogap and its effects on the single
particle properties of the system.

Considerable experimental evidence has accumulated which suggests that the
microscopic pairing mechanism of high-Tc \SC may be already manifested at a 
relatively short length scale. Both \SCg coherence and spin-spin correlation 
lengths in high-Tc cuprates are relatively short. The in-plane \SCg coherence 
length, for example, is estimated to be roughly $\sim 15\AA$ which indicates 
that Cooper pairs, on average, span only a few lattice spacings. Also, the 
correlation length for spin degrees of freedom estimated from a relatively broad 
resonance peak width in a momentum space is also of the order of a few lattice 
spacings \cite{bourges1,bourges2,bourges3,dai1,fong}. In addition, STM measurements of
the influence of magnetic impurities on local electronic structure of
high-Tc cuprates clearly suggest that Cooper pairs are local entities, and
can exist in a microscopically confined region \cite{hudson1}. 

These measurements suggest that a theoretical description based on a real space 
representation(in contrast to the BCS  momentum-space representation) in which 
fluctuations in a relatively localized region are rigorously treated can be a 
reasonable starting point to study the \SCg instability. 

In this paper we show that a cluster-based dynamical
mean-field theory approach is particularly useful for this purpose. Based on 
this approach, we are able to study the nature of short-ranged
quantum fluctuations of a strongly correlated system in the presence of well-defined \SCg and \AFc broken symmetries.\\ 

\indent The numerical results we report in this paper are an extension of 
earlier work\cite{watanabe2}
(unpublished) and are based on a Hubbard-like 
lattice model consisting of
two-site clusters using the self-consistent dynamical mean-field theory
approach of Kotliar and Georges\cite{georges1}. In the dynamical
mean-field theory approach, a lattice problem (for which an exact solution
is generally difficult to obtain) is mapped onto an impurity problem
coupled to an effective bath (which is typically much more tractable) -
thus the advantage of this theory is that once the model is constructed
it can be solved in principle exactly without further approximations when 
supplemented by a self-consistency condition which is derived by requiring
that the mean-field theory becomes exact in $d \rightarrow \infty$ limit
(d: dimension)\cite{vollhardt1}. The crucial point here is that the
effective bath is allowed to be time-dependent and thus highly non-trivial
quantum nature of (local) dynamics is retained. This approach has offered
new insights to the physics of strongly correlated electron systems which
are difficult to obtain from pre-existing theories \cite{georges2}. \\

\indent Here, in order to address the quantum aspects of {\it non-local}
but short-ranged fluctuations, we extend the original formalism to a
self-consistent two-site cluster in order to be able to explore the \SCg part of 
the phase diagram. In order to treat \SC, we
explicitly allow a $U(1)$ gauge symmetry breaking for both
diagonal(on-site) \cite{georges3} and off-diagonal(nearest-neighbor) 
\cite{LK} pairing channels inside a cluster. Within this model we are able to 
confirm the presence of a \SCg phase for reasonable Hubbard-type parameters. By 
virtue of enforcing self-consistency, our model effectively mimics an infinite
system, and thus is able to sustain a generic symmetry breaking. Thus, our
model facilitates a study of short-range fluctuations in a well-defined
broken symmetry phase such as \SCg or N$\acute{e}$el order. 

Within this generalized model we are able to show that the onset of \SC is 
intimately coupled to \AFc spin
fluctuations and that a resonance-like feature is indeed seen to emerge in the
dynamical spin susceptibility as the \SCg order parameter switches on.
Because we are able to solve the local cluster problem essentially exactly, the presence
of the Hubbard on-site repulsion $U$ will automatically favor an off-diagonal (nearest neighbor)
\SCg order parameter over an on-site one. We interpret this by analogy with BCS mean field
theories of the cuprate superconductors \cite{doniach2, dickinson} as favoring a d-wave like as opposed to an s-wave like order parameter symmetry.  Recently Lichtenstein and Katsnelson
\cite{LK} and Maier et al
\cite{maier} have developed approaches to dynamical cluster calculations for clusters with 4 or more sites
which show properties 
similar to those found in our coupled 2-site model. Our calculations allow us to 
examine the spectra of excitations in our representation of the strongly 
correlated system which tie in well with experimental observations. The approaches
of Lichtenstein and Katsnelson\cite{LK} and of Maier et al\cite{maier}
 may be expected to yield similar results. Recently, several groups \cite{chubukov,palee}
have argued that the resonance peak seen in inelastic neutron scattering may be thought of as a
type of ``spin - flip exciton". Our results suggest that the resonance is a generic property of
the strongly coupled superconducting state and is not dependent on details of the cuprate band
structure. \\ 

\indent The following is the organization of
the paper. Section II will describe the details of our methodology and
technical aspects. In section III, we present our results as well as the
interpretation. In section IV, we discuss the overall aspects of the
results obtained in section III and also some issues which were not
addressed in the preceding sections. We then suggest some future work and
close the section with a summary. Supplemental derivations and
discussions are given in appendices.\\\\

{\bf II. Methodology}

{\bf A. Model and Formalism} \\\\ \indent A two-site cluster model is the
minimal model needed to address the nonlocal nature of systems with 
order parameters for \AF (staggered magnetization) and \SC
(nearest-neighbor pairing) and we will base our analysis on this
model (Fig.1). We retain on-site Coulomb repulsion $U$, intra-cluster
hopping $t^{\prime}$ and inter-cluster hopping $t$. (The reason we
distinguish inter-cluster from intra-cluster hopping will become clear
below.) Thus, the physics which our model represents is described by the
following partition function:\\ \begin{eqnarray} Z=&Tr& [e^{-\beta
H}]\nonumber\\ \nonumber\\
H=&-&t\sum_{<i,j>,\sigma}\hat{d_{A}}_{i\sigma}^{\dagger}\hat{d_{B}}_{j\sigma}-t^{\prime}\sum_{(i,j),\sigma}\hat{d_{A}}_{i\sigma}^{\dagger}\hat{d_{B}}_{j\sigma}\nonumber\\
&+&(\epsilon_{d}-\mu)\sum_{i,\sigma}\hat{n_{i\sigma}}\nonumber\\
&+&U\sum_{i}\hat{n_{A}}_{i\uparrow}\hat{n_{A}}_{i\downarrow}+U\sum_{j}\hat{n_{B}}_{j\uparrow}\hat{n_{B}}_{j\downarrow}\nonumber\\
&-&h_{z}\sum_{i}(\hat{n_{A}}_{i\uparrow}-\hat{n_{A}}_{i\downarrow})+h_{z}\sum_{j}(\hat{n_{B}}_{j\uparrow}-\hat{n_{B}}_{j\downarrow})\nonumber\\
&+&\eta\sum_{<i,j>,\sigma}\hat{d_{A}}_{i\sigma}^{\dagger}\hat{d_{B}}_{j-\sigma}^{\dagger}\\
&+& (h.\; c.)  \end{eqnarray} where $\epsilon_{d}$ is a bare energy level
of the d-orbital of our model (i.e., $Cu-3d_{x^{2}-y^{2}}$ for
high-Tc cuprates) and $\mu$ is a chemical potential.
$\hat{d_{A}}_{i\sigma}^{\dagger}$ is the creation operator of an electron
with spin $\sigma$ on the $i$th site of the A sublattice.  $<i,j>$ and $(i,j)$
represent the sum over the inter and intra-cluster nearest neighbors,
respectively. $h_{z}$ is an infinitesimal staggered magnetic field. $\eta$
is an infinitesimal U(1) gauge symmetry breaking field for off-diagonal
pairing, which we take to be real (and all the anomalous components as
well). \\

 \indent Now, we self-consistently embed our cluster
into a lattice. Although the details of energetics of charge and spin
degrees of freedom may depend on the lattice, most of the qualitative
physics which appear to be manifested in real systems are shared by the
tight-binding Hubbard-like lattice models embedded on a lattice which has
a bipartite nature and a relatively smooth, structureless DOS
\cite{rozenberg1,jarrel1,pruschke1}. In this study, we chose to work with
a Bethe lattice.\\ \indent The self-consistency condition basically arises
as a result of seeking a homogeneous solution in the lattice problem. 
Here, we essentially follow the prescription developed by Kotliar and
Georges \cite{georges2,kotliar1} and extend it to the case of two-site
clusters in the presence of \SC. The basic idea is to systematically
expand (1) with respect to the {\it inter-cluster} hopping and pairing
amplitude (scaled as $t = \frac{t^{*}}{\sqrt{d}}, \:\eta =
\frac{\eta^{*}}{\sqrt{d}}$) and integrate out over the ligand degrees of
freedom. Due to the above scaling, to order $O(1)$, only the lowest order
term(the two-point propagator of the ligand multiplied by the probing
fields) survives the integration and we obtain the following mean-field partition function\\
\begin{equation} Z_{MF}=\int_{local}D\hat{\psi}D\hat{\psi}^{\dagger}
e^{-S_{eff}} \end{equation} where \begin{eqnarray*}
S_{eff}=&&\int_{0}^{\beta}\int_{0}^{\beta}d\tau
d\tau^{\prime}\hat{\psi}^{\dagger}(\tau)\hat{G}_{o}^{-1}(\tau -
\tau^{\prime})\hat{\psi}(\tau^{\prime})\\
&-&U\int_{0}^{\beta}d\tau(\hat{n}_{A\uparrow}(\tau)-\frac{1}{2})(\hat{n}_{A\downarrow}(\tau)-\frac{1}{2})\\
&-&U\int_{0}^{\beta}d\tau(\hat{n}_{B\uparrow}(\tau)-\frac{1}{2})(\hat{n}_{B\downarrow}(\tau)-\frac{1}{2})
\end{eqnarray*} $\hat{\psi}^{\dagger} = (\hat{d_{A}}_{\uparrow}^{\dagger},
\hat{d_{B}}_{\uparrow}^{\dagger},\hat{d_{A}}_{\downarrow},
\hat{d_{B}}_{\downarrow})$ is a four-component Nambu-Gorkov spinor
representing the local degrees of freedom.  $\hat{G}_{o}$ is a $4 \times 4$
cavity propagator for the cluster which can be thought of as a local
noninteracting propagator of a particular cluster where only the on-site U
for this cluster is turned off but all the other U are still present. From
the above effective local action , a total propagator $G(\tau)$ can be
calculated as:\\ \begin{equation} \hat{G}(\tau -
\tau^{\prime})=\frac{\int_{local}D\hat{\psi}D\hat{\psi}^{\dagger}\hat{\psi}(\tau)\hat{\psi}^{\dagger}(\tau^{\prime})e^{-S_{eff}}}{Z_{MF}}
\end{equation} Once a total propagator is obtained, the cavity propagator in the Bethe lattice case is
then given through the following self-consistency condition which can be
conveniently expressed in frequency space:\\ 
\begin{eqnarray*} G_{o}^{-1}(i\omega_{n})&=&\left( \begin{array}{c}{\cal
D}_{AA}^{\uparrow\uparrow}(i\omega_{n}) \; {\cal
D}_{AB}^{\uparrow\uparrow}(i\omega_{n}) \; {\cal
A}_{AA}^{\uparrow\downarrow}(i\omega_{n}) \; {\cal
A}_{AB}^{\uparrow\downarrow}(i\omega_{n})\\\\ {\cal
D}_{BA}^{\uparrow\uparrow}(i\omega_{n}) \; {\cal
D}_{BB}^{\uparrow\uparrow}(i\omega_{n}) \; {\cal
A}_{BA}^{\uparrow\downarrow}(i\omega_{n}) \; {\cal
A}_{BB}^{\uparrow\downarrow}(i\omega_{n})\\\\ {\cal
A}_{AA}^{\downarrow\uparrow}(i\omega_{n}) \; {\cal
A}_{AB}^{\downarrow\uparrow}(i\omega_{n}) \; {\cal
D}_{AA}^{\downarrow\downarrow}(i\omega_{n}) \; {\cal
D}_{AB}^{\downarrow\downarrow}(i\omega_{n})\\\\ {\cal
A}_{BA}^{\downarrow\uparrow}(i\omega_{n}) \; {\cal
A}_{BB}^{\downarrow\uparrow}(i\omega_{n}) \; {\cal
D}_{BA}^{\downarrow\downarrow}(i\omega_{n}) \; {\cal
D}_{BB}^{\downarrow\downarrow}(i\omega_{n})\\
\end{array} \right)  \end{eqnarray*}\\ 
where \begin{eqnarray*}
{\cal D}_{AA}^{\uparrow\uparrow}(i\omega_{n})&=&i\omega_{n}-\epsilon_{d}+\mu-\frac{U}{2}-
t^{2}G_{BB}^{\uparrow\uparrow}(i\omega_{n}),\nonumber\\
{\cal A}_{AA}^{\uparrow\downarrow}(i\omega_{n})&=&t^{2}F_{BB}^{\uparrow\downarrow}(i\omega_{n}),\nonumber\\
{\cal D}_{AB}^{\uparrow\uparrow}(i\omega_{n})&=&t^{\prime}-t^{2}G_{BA}^{\uparrow\uparrow}(i\omega_{n}),\nonumber\\
 \end{eqnarray*}  and similarly for the other elements of $ G_{o}^{-1}$. Full expressions are given in the Appendix.\\

\indent A few comments are in order. As a result of the decomposition of the
lattice into clusters and treating the physics inside and outside the
clusters in a different fashion (see, however, \cite{schiller1,georges6}),
the ratio of the bare values of the inter- and intra-cluster hopping matrix elements
no longer reflects the actual physics so that they should not be
compared  directly. Here, our main focus is on the  short-range \SCg and \AFc spin
fluctuations; i.e. that at part of the 
physics which intrinsically depends on the short-range dynamics, should
not depend  qualitatively on the construction  of the
lattice, and should still be manifested in the intra-cluster dynamics.
Considering the overall qualitative consistency of our results with
experiments (please see Sec.III), we believe that the physics presented here
is not an artifact due to the specific construction of our model.\\

\indent As will be shown below (see results) the self consistent numerical solution of the above model leads to a 
phase diagram exhibiting both antiferromagnetism and superconductivity. However, we find that the superconducting critical temperature is
relatively low for us to be able to obtain reliable numerical results for the excitation spectrum of the model. In order to strengthen the
tendency of the model to go \SCg, we perform a heuristic extension of the cluster expansion of the lattice action to higher orders in $\frac{1}{d}$
and incorporate the effect of inter-cluster particle-particle correlations. For example, at $O(\frac{1}{d})$, by employing the idea of the Hartree-Fock theory, we can extract a inter-cluster particle-particle correlation from the following four-point propagators in a local effective action $S_{eff}$:\\
 \begin{eqnarray*}
\int_{0}^{\beta}\int_{0}^{\beta}\int_{0}^{\beta}\int_{0}^{\beta}d\tau_{1}d\tau_{2}d\tau_{3}d\tau_{4}\hat{d_{oA\uparrow}}
(\tau_{1})^{\dagger}\hat{d_{oB\downarrow}}(\tau_{2})^{\dagger}\hat{d_{oA\downarrow}}(\tau_{3})\hat{d_{oB\uparrow}}(\tau_{4})\hat{G}_{BABA}^
{\uparrow\downarrow\downarrow\uparrow}(\tau_{1},\tau_{2};\tau_{3},\tau_{4})\nonumber\\\nonumber\\
\sim|F(0)|^{2}\int_{0}^{\beta}\int_{0}^{\beta}d\tau_{1}d\tau_{2}
\hat{d_{oA\downarrow}}^{\dagger}(\tau_{1})\hat{d_{oB\downarrow}}
(\tau_{2})\hat{d_{B\uparrow}}^{\dagger}(\tau_{1})\hat{d_{A\uparrow}}
(\tau_{2})\nonumber\\
+|F(0)|^{2}\int_{0}^{\beta}\int_{0}^{\beta}
d\tau_{1}d\tau_{2}\hat{d_{oB\downarrow}}^{\dagger}
(\tau_{1})\hat{d_{oA\downarrow}}(\tau_{2})
\hat{d_{A\uparrow}}^{\dagger}(\tau_{1})\hat{d_{B\uparrow}}(\tau_{2})\nonumber 
\end{eqnarray*}
where $|F(0)|$ is the average of equal-time inter-cluster anomalous Green's function. $\hat{d_{oA\sigma}}^{\dagger}$ and
$\hat{d_{A\sigma}}^{\dagger}$ represent local and neighboring cluster degrees of
freedom respectively. In the above, since each local degree of freedom
carries a pre-factor $\frac{1}{\sqrt{d}}$ and we sum it over all $d$ nearest neighbors, the above term becomes the order of $O(\frac{1}{d})$.\\
\indent We then incorporate the inter-cluster pairing correlation from higher orders in $\frac{1}{d}$ by employing the idea of the BCS theory. In order to preserve the effects of translational invariance within the Bethe lattice model, we then substitute the inter-cluster pairing by the intra-cluster pairing correlations which can be rigorously calculated :\\
 \begin{eqnarray*}
&&-g^{inter}\sum_{(i,j),\sigma}\hat{d_{oA}}_{i\sigma}^{\dagger}\hat{d_{B}}_{j-\sigma}^{\dagger}\hat{d_{B}}_{j-\sigma}\hat{d_{oA}}_{i\sigma}\nonumber\\
&\simeq&-g^{inter}\sum_{(i,j),\sigma}<\hat{d_{oA}}_{i\sigma}^{\dagger}\hat{d_{B}}_{j-\sigma}^{\dagger}>\hat{d_{B}}_{j-\sigma}\hat{d_{oA}}_{i\sigma}+
g^{inter}\sum_{(i,j),\sigma}\hat{d_{oA}}_{i\sigma}^{\dagger}\hat{d_{B}}_{j-\sigma}^{\dagger}<\hat{d_{B}}_{j
-\sigma}\hat{d_{oA}}_{i\sigma}>\nonumber\\
&\simeq&-g^{inter}\sum_{(i,j),\sigma}<\hat{d_{oA}}_{i\sigma}^{\dagger}\hat{d_{oB}}_{j-\sigma}^{\dagger}>\hat{d_{B}}_{j-\sigma}\hat{d_{A}}_{i\sigma}+
g^{inter}\sum_{(i,j),\sigma}\hat{d_{A}}_{i\sigma}^{\dagger}\hat{d_{B}}_{j-\sigma}^{\dagger}<\hat{d_{oB}}_{j-\sigma}\hat{d_{oA}}_{i\sigma}>\nonumber\\
&\equiv&-g^{inter}\left(\sum_{(i,j),\sigma}\xi_{AB}^{\sigma\:-\sigma}\hat{d_{B}}_{j-\sigma}\hat{d_{A}}_{i\sigma}
+\sum_{(i,j),\sigma}\hat{d_{A}}_{i\sigma}^{\dagger}\hat{d_{B}}_{j-\sigma}^{\dagger}\xi_{BA}^{-\sigma\:\sigma}\right)\nonumber
\end{eqnarray*}
 where $<\hat{d_{oA}}_{i\sigma}^{\dagger}\hat{d_{oB}}_{j-\sigma}^{\dagger}>\equiv \xi_{AB}$ 
is the thermal average of instantaneous intra-cluster
pairing which is calculated from the local quantum dynamics. 
For simplicity inter-cluster and intra-cluster coupling constants are taken to be the
same, $g^{intra} = g^{inter}$ and we set their value equal to $t$. We then take a mean-field limit with
respect to the inter-cluster hopping matrix elements as well, i.e., $\xi =
\frac{\xi^{*}}{\sqrt{d}}$, and obtain the following equations: \pagebreak
\begin{eqnarray} {\cal D}_{AA}^{\uparrow\uparrow}(i\omega_{n})&=&i\omega_{n}-\epsilon_{d}+\mu+h_{z}-\frac{U}{2}-
t^{2}G_{BB}^{\uparrow\uparrow}(i\omega_{n})\nonumber\\
&-&t^2\xi_{AB}^{\uparrow\downarrow}\xi_{BA}^{\downarrow\uparrow}G_{BB}^{\downarrow\downarrow}(i\omega_{n})\nonumber\\
&-&t^2(\xi_{AB}^{\uparrow\downarrow}F_{BB}^{\downarrow\uparrow}(i\omega_{n})+\xi_{BA}^{\downarrow\uparrow}F_{BB}^{\uparrow\downarrow}(i\omega_{n})),\nonumber\\
\\
{\cal A}_{AA}^{\uparrow\downarrow}(i\omega_{n})&=&t^{2}F_{BB}^{\uparrow\downarrow}(i\omega_{n})\nonumber\\
&-&t^2\xi_{AB}^{\uparrow\downarrow}\xi_{BA}^{\uparrow\downarrow}F_{BB}^{\downarrow\uparrow}(i\omega_{n})\nonumber\\
&+&t^2(\xi_{AB}^{\uparrow\downarrow}G_{BB}^{\downarrow\downarrow}(i\omega_{n})-\xi_{BA}^{\uparrow\downarrow}G_{BB}^{\uparrow\uparrow}(i\omega_{n})),\nonumber\\
&etc&\nonumber\\\end{eqnarray} where only representative components are shown
and the full expressions are given in the Appendix. 

We will see
that the effects of the additional inter-cluster coupling are manifested in all the
components of the propagator. As we show in section III, \SC already appears
in the absence of inter-cluster pairing correlation terms. When the additional
inter-cluster coupling is included we find as expected that the inter-cluster
coupling stabilizes the
\SCg phase. In particular, $T_{c}$ is enhanced for all the doping levels
and the single-particle gap (i.e., the superconducting gap) becomes more pronounced. This enhancement
in turn enables us to perform computations 
at higher temperature and hence to obtain more detailed excitation spectra in the
superconducting region of the phase diagram.\\\\

 {\bf B. Numerics} \\\\ \indent The local dynamics (3) and
(4) is basically that of a degenerate impurity Anderson model(with the
U(1) gauge symmetry breaking) and can be solved in various ways. Here, we
chose to solve it numerically using the quantum monte carlo(QMC) algorithm
of Hirsch and Fye \cite{hirsch1}. Since there are two vertices in the
local action (3), two Ising variables are introduced in our case. Then
importance sampling is performed sequentially based on the ratio of the
statistical weights for flipping Ising variables, which are given by
\cite{watanabe2}:\\ \begin{eqnarray*}
R_{A}(\tau_{i})=((e^{\lambda\sigma_{A}(\tau_{i})}-e^{-\lambda\sigma_{A}(\tau_{i})})G_{AA}^{\uparrow\uparrow}(\tau_{i},\tau_{i})+e^{-\lambda\sigma_{A}(\tau_{i})})\\
\;\;\;
\times((e^{\lambda\sigma_{A}(\tau_{i})}-e^{-\lambda\sigma_{A}(\tau_{i})})G_{AA}^{\downarrow\downarrow}(\tau_{i},\tau_{i})+e^{-\lambda\sigma_{A}(\tau_{i})})\\
-(e^{\lambda\sigma_{A}(\tau_{i})}-e^{-\lambda\sigma_{A}(\tau_{i})})^{2}F_{AA}^{\uparrow\downarrow}(\tau_{i},\tau_{i})F_{AA}^{\downarrow\uparrow}(\tau_{i},\tau_{i})\\\\
R_{B}(\tau_{i})=((e^{\lambda\sigma_{B}(\tau_{i})}-e^{-\lambda\sigma_{B}(\tau_{i})})G_{BB}^{\uparrow\uparrow}(\tau_{i},\tau_{i})+e^{-\lambda\sigma_{B}(\tau_{i})})\\
\;\;\;
\times((e^{\lambda\sigma_{B}(\tau_{i})}-e^{-\lambda\sigma_{B}(\tau_{i})}) 
G_{BB}^{\downarrow\downarrow}(\tau_{i},\tau_{i})+e^{-\lambda\sigma_{B}(\tau_{i})})\\
-(e^{\lambda\sigma_{B}(\tau_{i})}-e^{-\lambda\sigma_{B}(\tau_{i})})^{2}F_{BB}^{\uparrow\downarrow}(\tau_{i},\tau_{i})F_{BB}^{\downarrow\uparrow}(\tau_{i},\tau_{i}) 
\end{eqnarray*} where $\sigma_{A(B)}(\tau_{i})$ is the Ising variable at a
time step $\tau_{i}$({\it i}=1,...,L; L:the number of time slices) which
is introduced to decouple the Hubbard U term of A(B) site in a cluster.
Here, we would like to warn the reader not to confuse Ising variables
with electron spins. Ising variables are basically the fluctuating
fictitious magnetic field which couple to electron spins. $\lambda$ is a
constant which is given by the relation, $cosh(\lambda)=e^{\frac{\beta
U}{2L}}$. $G_{AA}^{\uparrow\uparrow}(\tau_{i},\tau_{i})$ is a diagonal
component(in terms of imaginary time index) of the instantaneous Matsubara
Green's function {\it at time step $\tau_{i}$ of an arbitrary sweep}, and
should not be confused with the averaged Matsubara Green's function
defined in (4). If the flip at the time slice $\tau_{k}$ is accepted, each
component of the Matsubara Green's functions is updated according to the
following rule:\\ \begin{eqnarray} \hat{G}^{new}_{ab}(\tau_{i},
\tau_{j})=\hat{G}^{old}_{ab}(\tau_{i}, \tau_{j}) \quad \quad \quad \quad
\quad \quad \quad \quad \quad \quad \quad \quad \quad \nonumber\\
+\sum_{c}\frac{(\hat{G}^{old}_{ac}(\tau_{i},
\tau_{k})-\delta_{ac}\delta_{ik})(e^{\Delta
V_{cc}(\tau_{k})}-1)}{1+(1-\hat{G}^{old}_{cc}(\tau_{k},
\tau_{k}))(e^{\Delta V_{cc}(\tau_{k})}-1)}\hat{G}^{old}_{cb}(\tau_{k},
\tau_{j})\nonumber\\ \end{eqnarray} \begin{eqnarray} \Delta
V_{ab}(\tau_{i})&=&-2\lambda\sigma_{A}(\tau_{i})|A\uparrow><A\uparrow|\nonumber\\
& &-2\lambda\sigma_{A}(\tau_{i})|A\downarrow><A\downarrow| \nonumber\\ &
&-2\lambda\sigma_{B}(\tau_{i})|B\uparrow><B\uparrow|\nonumber\\ &
&-2\lambda\sigma_{B}(\tau_{i})|B\downarrow><B\downarrow|\nonumber\\
\nonumber\\ a,b,c &=& \{A\uparrow, \; B\uparrow, \; A\downarrow, \; 
B\downarrow\} \nonumber\\ i,j,k &=& 1, .... , L \nonumber \end{eqnarray}
where $\Delta V_{ab}(\tau_{i})$ is a diagonal tensor which represents the
amount of change in the action associated with the flip of an Ising
variable at time slice $\tau_{i}$.\\ \indent The above set of equations
(3)-(5) is iterated until the convergence is achieved.  In practice, a
particular form of starting propagator is guessed depending on the type of
the anticipated solution. The impurity problem (3) and (4) might be solved
more efficiently by combining exact diagonalization and a Pad$\acute{e}$
approximation \cite{georges2}. The energy resolution of this approach
basically depends on the number of energy levels to be used as a basis.
Thus, it may not be the best approach when the spectral weight is widely
spread and a tiny feature is under search (such as a {\it dip} structure
in the ARPES spectrum in this case. Please see Section 8.2). Probably,
this approach is best used when the shape of the spectral function is
qualitatively known or reasonably guessed.\\ \indent A few remarks are in
order regarding the sampling algorithm. A complete algorithm would be to
integrate over the entire phase space of $\sigma_{A}(\tau_{i})$ and
$\sigma_{B}(\tau_{j})(i,j = 1, 2,..., L_{A})$. The size of the entire
phase is $2^{L} \times 2^{L}$ and it will be impractical to sample all the
configurations for most of the temperatures studied. This can then be dealt
 with by importance sampling. 
However, with this algorithm, we encountered a negative sign problem at the lowest
temperature studied $(\beta = 96)$ for finite doping, and for higher
temperatures, although the SDW persists for finite doping, we found no sign of
development of a gap in the single-particle spectra, and no sign of the onset of
or spin gap or building up of coherence in a staggered spin susceptibility,
and no sign of the onset of \SC. This phase basically corresponds to an
antiferromagnetic metal. In the DMFT based formalism, a metal-insulator
transition is described as the onset of {\it local or short-ranged}
gapless single-particle excitations and in this case, the transition is
driven by carrier doping and accompanied by \AFc order.\cite{watanabe4}\\

\indent Since our primary goal is to investigate the relationship between 
short-ranged \AFc spin fluctuations and \SC, we took a slightly different
approach. The idea is to implement an algorithm such that once the system
starts to build up an AF correlation (indicated by the appearance of a SDW
as mentioned above), two Ising variables are biased to stay at staggered
configurations, namely $(\uparrow, \downarrow)$ or $(\downarrow,
\uparrow)$. In other words, we attempt to build a machinery to do {\it
algorithmic projection} onto low-lying states. One way to implement this
idea is to use an identical random number array for both the A and B
sublattices. More specifically, we first attempt a flip of
$\sigma_{A}(\tau_{i})$ and see if $R_{A}(\tau_{i}) > x_{i}$(random number)
to accept the flip and update. Subsequently, we do the same for the
B-site using the same random number $x_{i}$. The motivation for this
algorithm is basically a sublattice symmetry of our model which is
manifested in the expressions for the accept ratio (please see the appendix for
more details), i.e., 
\begin{equation} R_{A}(\tau_{i}) =
R_{B}(\tau_{i})|_{\sigma_{B} = -\sigma_{A}, \: A \Leftrightarrow B, \:
\uparrow \Leftrightarrow \downarrow} \nonumber \end{equation} 
Once the
system passes into an antiferromagnetically correlated critical domain (by
which we mean a domain where the system acquires a tendency to pick
staggered configurations), the flip which causes a transition from a
parallel to a staggered configuration starts to get accepted more
frequently than the flip which causes the opposite transition. For the
sake of argument, let us consider the idealized case in which the
former flip is always accepted (that is to say, accept ratio for the
former flip is always greater than 1. Although this is generally not true
due to the presence of fluctuations, the argument below can be easily
generalized to a more realistic case).\\ 

\indent Suppose the Ising variables
for A and B sublattices are initially in the  $(\sigma_{A}, \sigma_{B}) =
(\uparrow, \downarrow)$ configuration at $\tau = \tau_{i}$. Then, due to the 
sublattice symmetry, if the flip of $\sigma_{A}$ is not accepted, the flip
of $\sigma_{B}$ is not likely to be accepted either, i.e., if
$R_{A}(\tau_{i}) < x_{i}$ for the flip $(\sigma_{A},
\sigma_{B}):(\uparrow, \downarrow) \Rightarrow (\downarrow, \downarrow)$,
then, it is likely that $R_{B}(\tau_{i}) < x_{i}$ for the flip $(\sigma
_{A}, \sigma_{B}):(\uparrow, \downarrow) \Rightarrow (\uparrow, \uparrow)$
as well due to the  sublattice symmetry. That is, if the initial
configuration is a staggered one, a conditional probability that a
$\sigma_{B}$ flip is not accepted given that a $\sigma_{A}$ is not
accepted is high. This, of course, depends on the level of noise, thermal
or quantum, already present, that is, the above statement holds in an
averaged sense. The better a sublattice symmetry is preserved at each time
slice, the stronger this correlation will be. Please note that if we use
a random number $y_{i}$ for the determination of a flip of $\sigma_{B}$
which is completely uncorrelated with $x_{i}$, then the probability of
a $\sigma_{B}$ flip solely depends on the value of $R_{B}(\tau_{ i})$ and a
rejection of $\sigma_{A}$ flip has no bearing on the $\sigma_{B}$ flip. If
the flip of $\sigma_{A}$ is accepted and the system is updated to
$(\downarrow, \downarrow)$ configuration, due to the presence of
antiferromagnetic correlation, the fli p of $\sigma_{B}$ will always be
accepted(as assumed above). Therefore, in this case, whether or not to use
the same random number will have no influence on the statistics.\\ \indent
If the initial configuration is a parallel one, say, $(\sigma_{A},
\sigma_{B}) = (\uparrow, \uparrow)$, then due to the presence of
antiferromagnetic correlation, a $\sigma_{A}$ flip will always be accepted
and the system is updated to $(\downarrow, \uparrow)$. Therefore, in this
case, it will make no difference whether or not to use the same random
number for the $\sigma_{B}$ flip. Thus, our algorithm is expected to bias
the system towards staggered configurations more than a standard
algorithm, when a system acquires an antiferromagnetic correlation. The
above argument can be easily generalized to the case where antiferromagnetic
correlations are not ideal, i.e., the probability of a flip from a parallel
to a staggered configuration is less than 1.\\ 

\indent Please note that, once
the system starts to lose antiferromagnetic correlations (by which we
mean that the transition probability between  staggered and parallel
configurations becomes equal), then the final population of the two
configurations will converge to the same value.  At this stage the effects of our
algorithm will also start to diminish, simply because  transitions from a
staggered to parallel configuration become more frequent and the
conditional probability discussed above becomes irrelevant. Therefore
our algorithm more or less approaches the  standard Metropolis-style algorithm when the
antiferromagnetic correlation diminishes, but starts to deviate from it
when the system passes into an antiferromagnetically correlated phase.
(Our algorithm respects  spin rotational symmetry in the paramagnetic
phase.) Please note that in our algorithm Ising variables can in principle
take all possible configurations and no symmetry on the state is
explicitly enforced. Therefore the symmetry and dynamics of the system still
depends on the physics and remain nontrivial.\\ 

\indent As will be seen from 
results in Sec.III, our algorithm leads to the onset of coherent \AFc spin
fluctuations for a certain parameter regime and unambiguously identifies a
close tie between the onset of SC and of a coherent \AFc spin fluctuation. 
We will see that the results obtained by our
algorithm, both for the dynamical and equilibrium properties of the model, are
qualitatively consistent with the experimental findings for high-Tc
cuprates. This suggests that the spin dynamics and its relation to the
onset of \SC as found with our algorithm is more or less consistent with
what is realized in real systems.\\ 

\indent Once the imaginary-time correlation functions are evaluated, equilibrium
quantities can be directly calculated from the self-consistent Matsubara
Green's functions, and dymanical quantities can be obtained from the
 via analytic continuation. We employed the Maximum Entropy(ME) \cite{silver1,gubernatis1} method to do
this. For each set of parameters, we perform the iteration (3)-(5) and (7) 
until all the main features in the dynamical spectrum converge. In this
study, we chose the number of time slices, $L$, to be such that
$\Delta\tau \equiv \frac{\beta}{L} = 0.25$ which should be sufficiently
small to obtain a reliable value for the physical quantities of interest.
\cite{hirsch2}.\\\\

{\bf III. Results and Discussion}\\\\

{\bf A. Choice of Parameters} \\\\ We choose $t=0.5$ throughout this paper
such that the bare bandwidth of the underlying Bethe lattice (equal to
$2t$) is the unit of the energy scale.  $U=2.2$ was chosen based on a
preliminary study on the one-band Hubbard model where we found
this makes the N\'eel temperature roughly optimal at half-filling.
Then
 we sampled a few $t^{\prime}$ values to search for the one which
 leads to a Neel state with the experimentally observed
$<S_{z}>$($\sim 0.5\mu_{B}$) at half filling. The optimal value for
$t^{\prime}$ appears to be located somewhere between 0.1 and 0.2. In this
paper, we show the results for
$t^{\prime}=0.2$. (Setting $t^{\prime}=0.1$ does not modify the results in
a qualitative way.) Although the above parameter setting has not been
carefully adjusted to be the optimal choice, the overall agreement
(qualitative and quantitative) with the experimental findings suggests
that the above set of values are reasonable to reflect those for real
materials. Again, let us emphasize here that our main interest is to
study the qualitative relationship between \SC and
antiferromagnetism.\\\\

{\bf B. Single-particle spectra} \\\\
 \indent We will present
single-particle spectra both with and without the inter-cluster pairing
coupling, and also compare them in the normal and \SCg phases.\\The
single-particle spectra are defined as follows:  \begin{eqnarray}
G(\omega)&=&-Im{Tr\hat{G}(i\omega)}\nonumber\\
&=&G_{AA}^{\uparrow\uparrow}(\omega)+G_{AA}^{\downarrow\downarrow}(\omega)+
G_{BB}^{\uparrow\uparrow}(\omega)+G_{BB}^{\downarrow\downarrow}(\omega) 
\end{eqnarray} Fig.2 shows the results at $\mu=0.6$($\delta \sim 0.05$) 
and $\mu=0.3$($\delta \sim 0.17$) {\it without} the inter-cluster pairing
coupling. Superconductivity sets in at $\beta \approx 16$ for
both dopings. The onset of \SC is determined by the onset of the average
of the {\it equal-time} anomalous Green's function which is defined as:
\begin{eqnarray*}
|F(0)|^{2}=|F_{AB}^{\uparrow\downarrow}(0)|^{2}+|F_{BA}^{\downarrow\uparrow}(0)|^{2}
+|F_{AB}^{\downarrow\uparrow}(0)|^{2}+|F_{BA}^{\uparrow\downarrow}(0)|^{2}
\end{eqnarray*} A single-particle gap starts to form when \SC sets in and
a quasiparticle-like feature appears on top of the valence band as the
temperature is lowered. Also, a {\it dip} which separates the
quasiparticle peak and the relatively broad feature develops.\\
\indent Next, we study the case {\it with} inter-cluster coupling included where we have assumed $g^{intra} = g^{inter}$.
As discussed above, this heuristic extension of our 2-site cluster model is put in to give
a more robust superconducting ground state, which allows us to extend our numerical calculations over a wider temperature range for studying the properties of the
superconducting state. 
Fig.3 shows a {\it normal} state single-particle spectrum for $\delta \sim 0.05 $ and $\beta = 16$ which was
obtained by setting all the anomalous components of the Green's function to zero(i.e., $\eta = 0$). We
confirmed that \SC indeed develops at this parameter setting when the anomalous components are turned on. As
we notice, in the absence of these components, the single-particle gap disappears and a sharp Kondo-like
peak develops at the Fermi level, indicating that holes which participate in the pairing are strongly
correlated.  Fig.4 shows the
the effect of inter-cluster pairing on the evolution of
single-particle Green's function as a function of temperature for
$\mu=0.7$($\delta \sim 0.04$) and $\mu=0.3$($\delta \sim 0.17$) where we 
have adopted the above values for $t'$ and $U$ for which $g \sim t$ as shown above.
Although the overall features are similar to Fig.2,
$T_{C}$ is increased(\SC sets in at $\beta \approx 10$) and the
single-particle gap becomes more robust especially at smaller doping. 
Experimentally, a single-particle gap has been found to decrease as a
function of doping \cite{shen4}.

\indent In all
the cases shown above, \SC is found to set in before a robust gap has been formed (In
particular, for $\mu = 0.3$ with inter-cluster pairing). As we will show
later, a signature of the onset of pairing fluctuations is more visible in
the pair-pair correlation function. The onset of pairing fluctuations
despite the absence of a robust gap is basically due to fluctuation
effects which are neglected in the BCS mean-field theory. Here we are able
to see this effect through our use of the dynamical mean-field theory approach.
In order to confirm that the gap formation is due to the onset of
\SC, we also studied the single-particle spectra in the normal state for
the same parameter settings. Fig.5 shows single-particle spectra at fixed
doping ($\delta \sim 0.04$) as a function of temperature in the normal
state. A ``Kondo peak" is manifested in the spectra\cite{jarrel1}. For
visual assistance, we overlay the results already shown in Fig.5(a) for the \SCg state. As the \SCg
amplitude grows, the spectral weight near $\epsilon_{F}$ is depleted and transfered to the
gap edges. These results unambiguously show that the single-particle gap is
driven by the onset of \SC. Fig.6 shows the single-particle
gap(determined by the FMHW), $T_{c}$, and $|F(0)|^{2}$(please see
Sec.III(D)) as a function of doping at $\beta = 16$. Except for the low doping region,
$\Delta_{1\:particle}$ roughly scales as $T_{c}$, which is what one would expect in the BCS
picture.\\ 

\indent Next we compare our results with the ARPES data. Since the single-particle spectra
calculated here are local, they should be in principle interpreted as {\it
angle-integrated} quantities. Thus, only a qualitative comparison can be
made with the angle resolved data. First, the quasiparticle peak as well
as the dip for both doping is qualitatively consistent with the ARPES
measurements. By comparing the spectra for the normal phase with those for
the \SCg phase, the appearance of the quasiparticle peak in the \SCg
phase can be qualitatively understood within the BCS picture in which
quasiparticles in the vicinity of $\epsilon_{F}$($|\epsilon-\epsilon_{F}|
\le \Delta_{1\:particle}$) in the normal phase form the Cooper pairs and
the quasiparticle peaks appear at both gap edges. The appearance of the
dip implies the separation of two energy scales, a sharp feature closer to
$\epsilon_{F}$ and a broad feature below the sharp one. This broad feature is
quite reminiscent of the ``hump" which has been observed in the ARPES measurements.
Recently, the microscopic origin of the ``hump" has been debated \cite{kivelson1, chubukov1},
and argued that the underlying band structure of high-Tc cuperates is important
for the onset of hump \cite{chubukov1}. Here, we would like to make some remarks
on this issue based on the results of our calculation. As shown above, the appearance
of the hump is strongly correlated with the onset of \SC (or a quasiparticle peak).
As we show later, a short-range \AFc spin fluctuation emerges as a coherent mode as
\SC sets in. (Please note that our calculation is essentially exact and considers
all energy scales involved). Therefore, it is reasonable to suspect that a hump-like
feature is due to  scattering of the photoelectrons by these coherent \AFc spin fluctuations.
In fact, the separation of the quasiparticle peak and the hump-like feature is roughly
proportional to the characteristic frequency of the \AFc spin fluctuations.
This is an indication that the appearance of the two energy scales is due to a resonance.
Please also note that since our calculation is fairly local, our results suggest
that a peak-dip-hump feature seen in ARPES measurements does not depend on the details
of the underlying band structure.\\
\indent The evolution of the single-particle gap as a
function of doping is also consistent the ARPES experiments
\cite{shen3,shen4,ding3} in which a pseudogap was found to monotonically
decrease in the underdoped region and fall off more rapidly in the
overdoped region as the doping increases. This suggest that  $T_{c}$
obtained in this study behaves more like $T^{*}$. This observation is also
consistent with the above finding that single-particle spectra start to
show an anomaly at $T_{c}$. The actual ARPES spectra show a more
pronounced quasiparticle peak than the one found here. This could be due
to the fact that our DOS is an angle-integrated quantity and is also
limited by the resolution of the Maximum Entropy algorithm.\\\\ 

{\bf C. 
Two-Particle Correlation functions} \\\\ \indent When a gap develops in a
single-particle channel, a coherence or rigidity usually start to build up
in two-particle channels. A natural question is then if and how
two-particle correlation functions show anomalous behavior as a result of
the onset of \SC. This question is also motivated by the fact that the
pseudogap has been observed in ARPES, ETM and STM (which basically measure
 single-particle spectra) and that INS, NMR and transport measurements
(which basically probe the two-particle correlation functions in spin and
charge sectors) have also found rich phenomena as the pseudogap sets in. 
In this section, we present local susceptibilities of relevant bosonic
degrees of freedom (both particle-hole and particle-particle channel) 
which appear to be playing an important role in high-Tc cuprates.  

{\bf
1. Spin degrees of freedom}\\\\
\indent Experimental studies of spin
dynamics have been one of the most valuable source of information on the
microscopics of high-Tc cuprates. In this section, we particularly focus
on resonance and a spin-gap features, and their relation to superconductivity. The onset of the
resonance which has been
observed in inelastic neutron scattering (INS) measurements as a narrow peak at the antiferromagnetic
wave vector
$\vec{Q}$=($\frac{\pi}{a}, \frac{\pi}{a}$)($a$:lattice parameter) and
energy $\sim 41meV$(near the optimal doping) \cite{rossat1, rossat2,rossat3,mook1}, seems to
be intimately related to the onset of \SC \cite{mook1, dai1}. The microscopic origin of this resonance
as well as its connection to the onset of \SC has been recently debated by a number of authors \cite{chubukov, palee, scz1}. The spin gap is characterized by a depletion of  low energy spectral weight in
the dynamical spin susceptibility at commensurate wave vector $\vec{Q}$
and its onset appears to be correlated with that of a single-particle gap
in ARPES spectrum \cite{dai1}. The relevant quantity is the dynamic staggered spin susceptibility
$\chi_{z}(\vec{Q},\omega)$ which basically contains the information about the collective excitations
of spin degrees of freedom at wave vector $\vec{Q}$. Using linear response theory,
$\chi_{z}(\vec{Q},\omega) $ is given as the following retarded staggered
spin-spin correlation function:\\ \begin{eqnarray*}
\chi_{z}(\vec{Q},\omega)&=-i&\int_{-\infty}^{\infty} dt e^{-i\omega
t}\theta(t)<[\hat{M}_{z}(t),\hat{M}_{z}(0)]> \end{eqnarray*} where the
staggered magnetization in our two-site cluster is defined as:
\begin{eqnarray*} \hat{M}_{z}(t)&=&\hat{S}^{A}_{z}(t)-\hat{S}^{B}_{z}(t)\\
&=&(\hat{n}_{A\uparrow}(t)-\hat{n}_{A\downarrow}(t))-(\hat{n}_{B\uparrow}(t)-\hat{n}_{B\downarrow}(t))\\
\end{eqnarray*} Then, by the fluctuation-dissipation theorem and analytic
continuation, $\chi_{z}(\vec{Q},\omega)$ can be directly calculated from
the imaginary-time staggered spin-spin correlation function which is given
as:  \begin{eqnarray*}
S_{z}(\vec{Q},\tau-\tau^{\prime})&=&<\hat{M}_{z}(\tau)\hat{M}_{z}(\tau^{\prime})>\\
&=&\frac{\int_{local}D\hat{\psi}D\hat{\psi}^{\dagger}\hat{M}_{z}(\tau)\hat{M}_{z}
(\tau^{\prime})e^{-S_{eff}}}{Z_{MF}}
\end{eqnarray*} The dynamic spin sucseptibility,
$\chi_{z}(\vec{Q},\omega)$ can then be calculated using the
fluctuation-dissipation theorem as follows:\\ \begin{eqnarray*}
\chi_{z}(\vec{Q},\omega)=(1-e^{-\beta\omega})S_{z}(\vec{Q},\omega) 
\end{eqnarray*} Please see the appendix for more details of the calculation.

\indent One remark should be made here.  $\chi_{z}(\vec{Q},\omega)$
defined above should be interpreted as a measure of short-range
fluctuations (i.e. the relevant length scale  represented by the clustering) due to
the two-site cluster nature of the model. That is, the above
$\chi_{z}(\vec{Q},\omega)$ is a response function defined for a probing
field which acts only on the two sites in our cluster instead of the
entire lattice. Thus, when N\'eel order sets in and the spin-spin
correlation length becomes sufficiently longer than the cluster size,
$\chi_{z}(\vec{Q},\omega)$ will start to deviate from the global staggered
susceptibility defined as a response function of the entire lattice. 
Although we will not have precise information on the correlation length
until we actually calculate it (probably by gradually increasing the
cluster size which will become numerically expensive as mentioned above
and is beyond the scope of this paper). Since a well-defined SDW is
severely degraded as soon as the system is doped, and as mentioned
earlier, the correlation length of spin fluctuations in high-Tc cuprates
away from half-filling as estimated by neutron scattering measurements is
roughly of the order of a few lattice spacing, we believe that the local
description given above should be a reasonable approximation.\\

\indent
Fig.7 and 8 show the evolution of $\chi_{z}(\vec{Q},\omega)$ as a
function of temperature for different doping levels with and without
inter-cluster pairing. In all cases, as the temperature is reduced, a
feature starts to grow at some energy scale and becomes more coherent at
lower temperatures. At the same time, low energy spectral weight is
gradually depleted and the spectrum eventually becomes gapped. These
features are quite reminiscent of the observed behavior of the resonance
and spin gap. Since the \SC starts to set in at $\beta \approx 10$ with
inter-cluster coupling and at $\beta \approx 16$ without it, they appear to
be correlated with the onset of \SC. The position of the resonance appears
to be rather insensitive to temperature variation. Next, we
plot $\chi_{z}(\vec{Q},\omega)$ as a function of doping at $\beta = 16$
for no inter-cluster coupling (Fig.9(a)) and with inter-cluster coupling(Fig9(b)).
 As long as \SC
persists, a peak basically remains visible. The position of the peak appears to increase
monotonically as a function of doping, and as a result, the spin gap also
increases as the doping is increased (at least in the underdoped region).
Upon passing into the overdoped region, the peak suddenly becomes
incoherent and the spin gap disappears. We also found that where a resonance remains coherent,
its energy is basically below a particle-hole continuum (i.e., $\omega_{res} \le 2\Delta_{sc}$.
See below for optical conductivity results). The overall qualitative feature of the
evolution of $\chi_{z}(\vec{Q},\omega)$ as a function of doping is
basically similar for both cases. 

Now we compare the above
results with the experiments. A resonance peak has been found by spin-flip
INS in the \SCg phase\cite{mook1,dai1,bourges2}. The position of resonance
$\omega_{res}$ has been observed to soften as the doping is decreased from
the optimal level \cite{bourges1,dai1,fong1} and appears to be relatively
temperature insensitive \cite{bourges2,dai1,fong2,bourges5} for a fixed
doping. The magnitude of a spin gap also appears to decrease as the doping
is reduced from optimal doping \cite{dai1, bourges2}. These features
are also qualitatively captured by our results. Furthermore, the relative
energy scales among the resonance, spin gap and single-particle gap as
found by INS and ARPES measurements are semi-quantitatively consistent
with the results shown in Fig.4, Fig.5, Fig.7 and Fig.8.\\ 

\indent To further
investigate the relation of the spin spectra with superconductivity, we also studied the
spectra in the {\it normal} phase. Fig.10 is $\chi_{z}(\vec{Q},\omega)$
for $\beta =16$ and $\delta \sim 0.17$.  For comparison, we also show the
data in the \SCg phase with roughly the same parameter settings. In the
normal phase, the coherence of the resonance is severely degraded and the
existence of the peak can no longer be recognized. Also,
$\chi_{z}(\vec{Q},\omega)$ acquires a substantial weight in the low energy
sectors and the spin gap is completely filled up. These findings imply that 
the short-range \AFc spin
fluctuations maintain their coherence by means of the pairing correlations.

 {\bf 2. Charge degrees of
freedom} \indent Next, we present the dynamics of charge degrees of
freedom. Since all the qualitative features found for the case of no
inter-cluster pairing are also found when the inter-cluster pairing is switched on, we
will only show the results for the latter case.
Optical measurements have also revealed the emergence of a gap feature at
some temperature $T^{*}$ which is higher than $T_{c}$ in the underdoped
region. A relevant physical quantity which probes the charge dynamics is
the optical conductivity which is directly related to the retarded
current-current correlation function. In our two-site cluster model, a
local current operator can be defined in the following form:\\
\begin{eqnarray*}
\hat{j}&=&-i\sum_{\sigma}({\hat{d}_{A\sigma}^{\dagger}(\hat{d}_{B\sigma}-\hat{d}_{A\sigma})-(\hat{d}_{B\sigma}^{\dagger}-\hat{d}_{A\sigma}^{\dagger})\hat{d}_{A\sigma}}) 
\end{eqnarray*} where the direction of current is that of intra-cluster
hopping. This is basically a lattice version of current operator which
takes a more familiar differential form in a continuum limit. Then, the
Kubo formula for the paramagnetic part of optical conduc tivity
$\sigma^{para}(\omega)$ is given in terms of the following retarded
current-current correlation function $\chi_{jj}(\omega)$:\\
\begin{eqnarray*} \chi_{jj}(\omega)&=&-i\int_{-\infty}^{\infty} dt
e^{-i\omega t}\theta(t)<[\hat{j}(t),\hat{j}(0)]>\\
\sigma^{para}(\omega)&=&i\frac{\chi_{jj}(\omega)}{\omega} \end{eqnarray*}

Since our clusters only have two sites, only the longitudinal part can be
defined. As in the case of spin dynamics, we first calculate the following
imaginary-time causal current-current correlation function:\\
\begin{eqnarray*}
X_{jj}(\tau-\tau^{\prime})&=&<\hat{j}(\tau)\hat{j}(\tau^{\prime})>\\
&=&\frac{\int_{local}D\hat{\psi}D\hat{\psi}^{\dagger}\hat{j}^{\alpha}(\tau)\hat{j}^{\alpha}(\tau^{\prime})e^{-S_{eff}}}{Z_{MF}}
\end{eqnarray*} Ther detailed  form of $X_{jj}(\tau-\tau^{\prime})$ is omitted
here \cite{watanabe2}. Then, $\sigma^{para}(\omega)$ can be obtained by
the fluctuation-dissipation theorem and analytic continuation:\\
\begin{eqnarray*}
\sigma^{para}(\omega)=-i\frac{1-e^{-\beta\omega}}{\omega}X_{jj}(\omega) 
\end{eqnarray*} We plot $Re[\sigma^{para}(\omega)]$ for $\mu=0.3$ and
$\mu=0.7$ in Fig.11. For $\mu=0.7$(Fig.11(a)), at higher temperatures, a
Drude-like peak is visible due to the fact that the single-particle
spectrum has a nonzero weight at $\epsilon_{F}$(Please see Fig.4(a)). As
the temperature is lowered, the Drude-like weight diminishes and
eventually a gap opens up as \SC develops. Even when the system passes
into the \SCg phase, since the single particle spectrum does not become
fully gapped, a small weight still remains in the low energy sectors
above $\beta = 16$. Again, this is due to fluctuation effects which are
not included in the BCS theory. The size of the gap is
quantitatively consistent with a single-particle gap (i.e.,
$\Delta_{optical} \simeq 2\Delta_{1\:particle}$). At this doping, a charge
gap feature (i.e., transition between the lower and upper Hubbard bands as
seen in Fig.11(a)) is visible for all temperatures. A feature
corresponding to the quasiparticle peaks in the SC state becomes visible
as the temperature is lowered($\beta = 24$). In Fig.11(b)(for $\mu=0.3$),
a large Drude-like peak appears at high temperatures due to a substantial
weight at $\epsilon_{F}$ in the single-particle spectrum(Fig.4(b)). As the
temperature decreases, the weight in the low energy sectors disappears and
a gap feature develops. Similarly, due to  residual single-particle
weight at $\epsilon_{F}$, a robust gap does not develop in charge spectra
until the temperature is lowered to $\beta=24$. Please note that a
feature for the charge gap for $\mu=0.3$ is hardly visible, which is
consistent with our single-particle spectra at this doping(Fig.4(b)). This
is basically due to a low electron density. An addtional feature on the
quasiparticle peak (i.e., the feature at the gap edge), however, is much
more robust at $\mu=0.3$ (as can be expected from Fig.4(b)) and starts to
show up already at $\beta=16$. This feature on the gap edge has been
observed in the infrared spectra for the optimally doped both single and
bi-layer high -Tc materials. The size of the gap for $\mu=0.3$ is reduced
from that for $\mu=0.7$ and is also quantitatively consistent with the
single-particle gap as shown in Fig.4(b). Please note the wide range of
energy scales over which the weight is redistributed as a function of
temperature. This strongly temperature-dependent spectral weight has been
observed in the optical conductivity spectrum of high-Tc cuperates
\cite{timusk1} and suggests that the charges participating in the pairing
are strongly correlated.\\ 

\indent What we showed (numerically) in the
above is basically that a paramagnetic current disappears in the \SCg
phase below a certain energy scale($\sim \Delta_{optical}$). Therefore,
this will naturally lead to the onset of the Meissner effect caused by a
residual diamagnetic response which comes from the mobile Cooper pairs. 
The magnitude of the diamagnetic response is proportional to the Cooper
pair density.\\ 

{\bf 3. Pair degrees of freedom} \indent We now turn to
the particle-particle channel. The quantity of interest is the pair-pair
correlation function. This function basically contains information on the
amplitude fluctuations of the \SCg order parameter. A local d-wave -- like
pair-pair correlation function $\chi_{d}(\omega)$ can be defined in the
following way:\\ \begin{eqnarray*}
\chi_{d}(\omega)&=&\int_{-\infty}^{\infty} dt e^{-i\omega
t}<\hat{\Delta}_{d}(t)\hat{\Delta}_{d}^{\dagger}(0)> \end{eqnarray*} where
a local d-wave order parameter centered at a particular A-site is
defined as:\\ \begin{eqnarray*}
\hat{\Delta}_{d}^{\dagger}(A,t)&=&\hat{d}_{A\uparrow}^{\dagger}(t)\hat{d}_{B1\downarrow}^{\dagger}(t)+\hat{d}_{A\uparrow}^{\dagger}(t)\hat{d}_{B2\downarrow}^{\dagger}(t)\\
&-&\hat{d}_{A\uparrow}^{\dagger}(t)\hat{d}_{B3\downarrow}^{\dagger}(t)-\hat{d}_{A\uparrow}^{\dagger}(t)\hat{d}_{B4\downarrow}^{\dagger}(t)\\
&-&\hat{d}_{A\downarrow}^{\dagger}(t)\hat{d}_{B1\uparrow}^{\dagger}(t)-\hat{d}_{A\downarrow}^{\dagger}(t)\hat{d}_{B2\uparrow}^{\dagger}(t)\\
&+&\hat{d}_{A\downarrow}^{\dagger}(t)\hat{d}_{B3\uparrow}^{\dagger}(t)+\hat{d}_{A\downarrow}^{\dagger}(t)\hat{d}_{B4\uparrow}^{\dagger}(t)
\end{eqnarray*} Please note that each pairing for a given
nearest-neighbours appears as a singlet which is a direct consequence of
taking a zero center-of-mass momentum of the pair(please see below). This
expression can be obtained by simply transforming back to real space
the following more familiar definition of d-wave operator in 2d momentum
space:\\ \begin{eqnarray*}
\hat{\Delta}_{d}^{\dagger}(t)=\sum_{\vec{p}=(p_{x},p_{y})}(cos(p_{x})-cos(p_{y}))\hat{d}_{\vec{p}\uparrow}^{\dagger}(t)\hat{d}_{-\vec{p}\downarrow}^{\dagger}(t) 
\end{eqnarray*} After performing an inverse Fourier transformation, this
leads to the following (global) d-wave -- like operator which is basically a sum
of the above local d-wave operator over the all A-sites:\\
\begin{eqnarray*}
\hat{\Delta}_{d}^{\dagger}(t)&=&\sum_{A}\hat{\Delta}_{d}^{\dagger}(A,t) 
\end{eqnarray*} The choice of A-site as a reference is arbitrary and
B-sites could be chosen just as well.\\ \indent Since the clusters only have 
two sites, we need to make an assumption about the symmetry of the
dynamics, namely a rotational symmetry among the four
nearest-neighborhoods. The above $\chi_{d}(\omega)$ is defined with
respect to a particular A-site and contains the pairing correlation
between its four nearest-neighboring B-sites. Here, we basically make the
assumption that the dynamics of these four pairings are the same except
for their phases. We view this as a way to represent the effects of inter-cluster correlations within
the constraints of our model\cite{LK}.  Therefore, the above local $\chi_{d}(A,\omega)$ can be
collapsed into a single two-site cluster. Then, we calculate its corresponding
imaginary-time causal correlation function:\\ \begin{eqnarray}
\chi_{d}(\tau-\tau^{\prime})&=&<\hat{\Delta}_{d}(\tau)\hat{\Delta}_{d}^{\dagger}(\tau^{\prime})>\nonumber\\
&=&4<F_{AB}^{\uparrow\downarrow\sigma(l)}(\tau,\tau)F_{BA}^{\downarrow\uparrow\sigma(l)}(\tau^{\prime},\tau^{\prime})>_{\sigma(l)}\nonumber\\
&+&4<F_{BA}^{\uparrow\downarrow\sigma(l)}(\tau,\tau)F_{AB}^{\downarrow\uparrow\sigma(l)}(\tau^{\prime},\tau^{\prime})>_{\sigma(l)}\nonumber\\
&+&4<F_{AB}^{\uparrow\downarrow\sigma(l)}(\tau,\tau)F_{AB}^{\downarrow\uparrow\sigma(l)}(\tau^{\prime},\tau^{\prime})>_{\sigma(l)}\nonumber\\
&+&4<F_{BA}^{\uparrow\downarrow\sigma(l)}(\tau,\tau)F_{BA}^{\downarrow\uparrow\sigma(l)}(\tau^{\prime},\tau^{\prime})>_{\sigma(l)}\nonumber\\
&-&4<G_{AB}^{\uparrow\uparrow\sigma(l)}(\tau,\tau^{\prime})G_{AB}^{\downarrow\downarrow\sigma(l)}(\tau^{\prime},\tau)>_{\sigma(l)}\nonumber\\
&-&4<G_{BA}^{\uparrow\uparrow\sigma(l)}(\tau,\tau^{\prime})G_{BA}^{\downarrow\downarrow\sigma(l)}(\tau^{\prime},\tau)>_{\sigma(l)}\nonumber\\
&-&<G_{AA}^{\uparrow\uparrow\sigma(l)}(\tau,\tau^{\prime})G_{BB}^{\downarrow\downarrow\sigma(l)}(\tau^{\prime},\tau)>_{\sigma(l)}\nonumber\\
&-&<G_{BB}^{\uparrow\uparrow\sigma(l)}(\tau,\tau^{\prime})G_{AA}^{\downarrow\downarrow\sigma(l)}(\tau^{\prime},\tau)>_{\sigma(l)}\nonumber\\
&+&\delta_{\tau\tau^{\prime}}<G_{AA}^{\uparrow\uparrow\sigma(l)}(\tau,\tau)+G_{BB}^{\uparrow\uparrow\sigma(l)}(\tau,\tau))>_{\sigma(l)}
\end{eqnarray} where $\chi_{d}(\tau-\tau^{\prime})$ is an average per
pair. Basically, the product of $F$'s and $G$'s correspond to a (local) 
pair hopping and pair breaking fluctuations, respectively. After the
analytic continuation (Maximum Entropy), we finally obtain a local
dynamical pair-pair correlation function $\chi_{d}(\omega)$.\\ \indent We
plot $\chi_{d}(\omega)$ for $\mu=0.7$(Fig.12(a)) and $\mu=0.3$(Fig.12(b)). 
We see that for both dopings, as the temperature is lowered, a narrow
feature grows around $\omega = 0$ which indicates that the \SCg order
parameter develops a well-defined static component. The temperature at
which this narrow feature develops basically corresponds to $T_{c}$
determined from the onset of instantaneous nearest-neighbor pairing
correlation.\\ \indent From a phenomenological point of view, pair hopping
and pair breaking fluctuations correspond to short-wavelength phase and
amplitude fluctuation of the \SCg order parameter, respectively. The latter is
more intuitively transparent but the former becomes readily clear if one
notes that:  \begin{eqnarray*} (pair \; hopping) 
&\sim&-t_{pair}\sum_{i,\:j}(\hat{\Delta}_{d,\:i}^{\dagger}\hat{\Delta}_{d,\:j}
+\hat{\Delta}_{d,\:j}\hat{\Delta}_{d,\:i}^{\dagger})\\
&\sim& -t_{pair}\rho_{sc} cos({\phi_{i}-\phi_{j}})  \end{eqnarray*} where
\begin{equation} <\hat{\Delta}_{d,\:i}^{\dagger}> \; = \;
<\hat{\Delta}_{d,\:i}> \; \sim \; \sqrt{\rho_{sc}} e^{i\phi_{i}}
\end{equation} \indent Therefore, we can see in more detail the nature of
the onset of a \SCg phase by studying both pair hopping ($F \times F$
terms in (15)) and pair breaking ($G \times G$ terms in (15))
fluctuations. From detailed studies \cite{watanabe6}, we found that
pair breaking fluctuations are more strongly gapped at $T_{c}$ for
$\mu=0.7$ than for $\mu=0.3$, and this implies that the amplitude
fluctuations are more enhanced for $\mu=0.3$ when \SC switches on. These
findings basically suggest that Cooper pairs are more tightly bound in
the low doping region (which is consistent with the doping evolution of
the single-particle gap). Since phase fluctuations are typically
energetically cheaper than an amplitude fluctuations (this will typically
be a gapless Goldstone mode in the long wavelength limit), a remedy for
the over-estimate of $T_{c}$ in the low doping region may be sought by
incorporating a phase fluctuation mechanism, or equivalently charge
localization in the dual picture. (In the present model, the coherent
pair-hopping fluctuations are gapless as seen in Fig.12, and  phase
coherence can not be suppressed, i.e., a charge localization mechanism is
absent.)  Although such localization mechanisms have been proposed
\cite{doniach1, ek1}, phenomenology in the lightly doped region is still
under debate\cite{chakravarty}. Please note that the physics which governs global
phase coherence is a low energy long-wavelength phenomenon (depending on the
localization length of Cooper pairs) and is thus not possible to address by
the present approach based on the small  cluster size.\\\\

 {\bf D. 
Equillibrium properties and  phase diagram} \\\\ \indent Since our main
focus is \AF and \SC, we investigate those two phases. We show our results
both with and without the
inter-cluster coupling.\\ \indent An \AFc phase is signaled by the onset of
the thermal average of a staggered magnetization induced by an
infinitesimal staggered field.  Here, instead, we chose to induce symmetry
breaking by appropriately initializing the Ising variables for each
sublattice. Although the staggered magnetization is in principle a vector
order parameter a natural choice is its z-component, since the
Hubbard-Stratonovich decomposition was performed along the z-axis. This is
given as:  \begin{eqnarray*}
<\hat{M}^{z}>&=&<\hat{S}_{A}^{z}>-<\hat{S}_{B}^{z}>\\
<\hat{S}_{A}^{z}>&=&<\hat{n}_{A\uparrow}>-<\hat{n}_{A\downarrow}>\\
<\hat{S}_{B}^{z}>&=&<\hat{n}_{B\uparrow}>-<\hat{n}_{B\downarrow}>
\end{eqnarray*} where $<...>$ means a thermal average. In Fig.13(a), the
staggered magnetization, $<\hat{M}^{z}>$, is plotted at half-filling as a
function of temperature. Since \SC does not set in exactly at half
filling, there is no distinction between the two cases.  The staggered
magnetization disappears roughly at $\beta = 6$. Upon doping,
$<\hat{M}^{z}>$ almost suddenly drops and disappears before $\delta$
reaches $\sim 0.009$.  A detailed study of the destruction of \AF as a
function of doping is outside the scope of this paper, and we did not
sample sufficiently many doping levels to determine the nature of
transition, such as the order of transition.\\ \indent Similarly, if we
employ the definition of the BCS picture, an onset of \SC can be defined
as the onset of an average of {\it equal-time} anomalous Green's function:
\begin{eqnarray*}
|F(0)|^{2}=|F_{AB}^{\uparrow\downarrow}(0)|^{2}+|F_{BA}^{\downarrow\uparrow}(0)|^{2}+
|F_{AB}^{\downarrow\uparrow}(0)|^{2}+|F_{BA}^{\uparrow\downarrow}(0)|^{2}
\end{eqnarray*} All the on-site pairing components,
$F_{AA}^{\uparrow\downarrow}(0)$, $F_{AA}^{\downarrow\uparrow}(0)$,
$F_{BB}^{\downarrow\uparrow}(0)$ and $F_{BB}^{\uparrow\downarrow}(0)$, are
always one to three orders of magnitude smaller. We turn on a small $\eta$
a nd see if the system can sustain a finite $|F(0)|^{2}$. Therefore,
according to  linear response theory, the appearance of a finite
$|F(0)|^{2}$ basically corresponds to divergence of the pairing fluctuations. 
In Fig.13, $|F(0)|^{2}$ is also plotted as a function of doping for
$\beta=16$(Fig.13(b)) and as a function of temperature for large
doping(Fig.13(c)) and small doping(Fig.13(d)). We found that the inclusion of inter-cluster coupling
systematically enhances $T_{c}$\cite{watanabe2}. The
doping which gives a maximum $|F(0)|^{2}$ seems to be somewhere between
$\mu=0.4$ ($\delta \sim 0.13$) and $\mu = 0.3$($\delta \sim 0.17$) for
both cases. A determination of the doping which gives a maximum $T_ {c}$
requires a detailed study involving varying both temperature and doping and was
difficult to identify with the same accuracy(it appears to be located at
least between $\delta \sim 0.009$ and $\delta \sim 0.17$).  Note that a
\SCg phase appears as soon as the system is doped.\\ \indent In Fig.14,
we show the phase diagram as a function of doping and temperature
in the presence of inter-cluster coupling. (A phase diagram without inter-cluster
pairing is also shown here. The overall topography is quite
similar\cite{watanabe2}.) Note that the overall topography is quite
consistent with that of high-Tc cuprates: a sudden suppression of AF order
away from half-filling, an upper critical doping for \SC at roughly
$\delta_{c} \sim 0.3$ and an optimal doping is located $0.13 \leq \delta
\leq 0.17 $. For $t^{\prime}=0.1$, all the qualitative features are
essentially the same although the upper critical doping and the staggered
magnetization at half-filling appear to increase slightly.\\ \indent One
notices, however, that a \SCg amplitude is quite alive even in a lightly
doped region, i.e., the lower critical doping is essentially zero. This
feature was found for a related model in a BCS mean-field approximation
\cite{doniach2}(which can be thought of as a static version of the present
approach) and later interpreted as the onset of a phase stiffness(i.e.,
the amplitude part) of the \SCg order parameter \cite{doniach1} which
exactly corresponds to $|F(0)|^{2}$ here.  A locally constructed \SC was
also found to coexist with \AF for a relatively wide range of doping in
the dynamical mean-field theory study of equilibrium state of a cluster
model in which local dynamics are more constrained \cite{LK}.  As
mentioned in the previous section, the physics of underdoped region is
still under debate.\\\\

{\bf IV. Summary and Conclusions.} \\\\

\indent As mentioned previously, the  relatively short coherence and spin-spin
correlation lengths which are observed in the in high-Tc cuprates constitute the key motivation of
the present study as well as underpinning the basis of our approach. The subject of
this paper is to investigate the role of short-range quantum fluctuations
in high-Tc superconductivity and, in particular, the interplay between pairing and spin
fluctuations. In order to achieve a more complete understanding of the 
strongly coupled dynamics whose
calculation goes beyond the bounds of BCS mean-field we apply dynamical
mean-field theory to a simple 2-atom cluster model. 
We showed that this model,
despite its simplicity, not only reproduces the main characteristic features of high-Tc
cuprates(i.e., the basic topography of the phase diagram, a quasiparticle
feature in single-particle spectra, the spin-gap, neutron resonance, d-wave-- like
pairing, etc) with qualitative and semi-quantitative consistency, but also
revealed a strong correlation between coherent short-ranged \AFc spin
fluctuations and pairing correlations: i.e., if \SC is suppressed the \AFc
spin fluctuations lose their coherence. Thus our approach allows us to directly study the
relation between the dynamics of the spin and pair degrees of freedom.

We would like to emphasize that the present model was
constructed from strongly correlated electron degrees of freedom only and
no prior assumption was made for the existence of any intermediate energy
scale which typically represents some sort of composite bosonic degrees of
freedom in either particle-hole or particle-particle channel. In the
present formalism, bosonic degrees of freedom arise {\it naturally} as a consequence of 
gap formation in
the single-particle sector. In our model, away from half filling, coherence builds up in a singlet
particle-particle channel and this is accompanied by a development of coherence in the staggered
spin channel at some characteristic energy scale. The only nontrivial
assumption that we put into the present model is basically that
the correlation length of spin and pairing fluctuations is sufficiently short
and that \AFc fluctuations are a relevant spin fluctuation mode to
consider. In this context we have been able to show that the spin fluctuation resonance which has been discussed by a
number of authors \cite{chubukov,palee} as a spin - flip exciton-like mode is a generic feature of our model. 
This suggests that it survives beyond the RPA approach and should not depend on details of the band structure of the
system. Thus it appears to be an intrinsic property of the strongly coupled superconducting state in the presence of
strong on-site repulsion.\\

 \indent From our results on the single particle excitations of the model, and the fact that we can correlate the ``hump
feature" seen in ARPES spectra with that seen in our model, we conclude that the spectral weight of the spin fluctuation
resonance is sufficient to show up as an energy loss peak in the photeoemmission spectra. This is in contrast to
recent arguments\cite{kivelson1,chubukov1} suggesting that details of the band structure are important in this
interpretation of the ARPES data. From the point of view of the local physics represented in our model, it appears that
this feature is an intrinsic property of Hubbard-type models of superconductivity.

 \indent The present approach is essentially an analogue
of Weiss mean-field theory in which fluctuations along the imaginary time
are exactly treated but all the spatial fluctuations whose wavelengths
are longer than the size of the cluster are averaged out. Unlike
theories which are based on momentum space representation, it
is based on a local picture. Therefore, when the physics in the
thermodynamic limit is dominated by low-lying long-wavelength modes,
this local description is expected to break down. (When the correlation length becomes
sufficiently longer than the cluster size but still finite, the present
model will incorrectly assume that true long-range order is established).
Typically, it is the dimensionality of the system which plays a crucial role in deciding
the fate of the thermodynamic limit when long-wavelength fluctuations
dominate. (Of course, the length scale of interactions among the local
order parameters, short-ranged or long-ranged, is also important and this
is assumed to be of short-range in the present study). Therefore, physics
which depends on the dimensionality in an essential way will be difficult
to address in this approach. (This approach is essentially a ``0 + 1''
dimensional formalism.) Instead, the present model is looking at the
portion of physics which is rather insensitive to the dimensionality, i.e.
the short-range short-wavelength modes. Such features are relatively
dimension independent -- indeed, \SC and similar anomalous features have
been reported in quasi-1D doped Hubbard ladder systems as well
\cite{uehara1}.\\

\indent Before we conclude, we note that recent studies \cite{LK, maier} have looked at
dynamical properties of four-site cluster models. Although these authors have not
yet investigated the excitation spectra of such models, we expect them to lead to
qualitatively the same conclusions that we have reached based on our 2-site cluster model.
A four-site square cluster model would 
also allow consideration of other kinds of long range order such as a chiral 
{\it flux phase} \cite{affleck1, wen1}. A four-site cluster model
would also allow for study of the dynamics of $\pi$-operators in the presence of
broken symmetry phases (effectively in a thermodynamic limit) and the
connection between SO(5) symmetry \cite{scz, scz1} and \SC. This would be 
complimentary to results of exact numerical work\cite{hanke1} for finite
clusters.  

\indent The present approach is to transform the original lattice problem to
an  effective self-consistent local which can be exactly solved in principle.
Thus, it can provide essentially exact solutions to the original problem within the
range of validity of this transformation. Certain aspects of high-Tc \SC
appear to fit reasonably well into the regime in which this transformation
is valid. The present model is perhaps the simplest nontrivial one in the
context of a self-consistent cluster model in the dynamical mean-field
theory approach. Yet, it appears to succeed in addressing some nontrivial
aspects of strongly correlated electron systems which would be difficult
to approach by other means. The results we have presented clearly suggest that
as long as the physics at hand is of short-range this approach can be
quite effective and promising.\\ 

\indent In summary, we constructed a
self-consistent two-site cluster model in the dynamical mean-field limit
in which short-range short-wavelength fluctuations of charge and spin
degrees of freedom are treated exactly in the presence of
superconductivity. The equilibrium properties of our model (\SC and
antiferromagnetism) as a function of doping and temperature reproduce the
overall qualitative and basic quantitative features of the phase diagram
of high-Tc cuprates. The behavior of single-particle and two-particle
spectra of our model in the \SCg phase can be interpreted to give a
consistent account of anomalous features of high-Tc cuprates such as the
pseudogap(charge and spin) and resonance observed in ARPES, INS, NMR and
other optical measurements. \\

{\bf ACKNOWLEDGEMENTS}\\\\We
acknowledge the NSF for support through Grant No. DMR9418964 and through
the Center for Materials Research at Stanford University, and the San Diego
Supercomputer Center for a grant of computer time.\\\\

{\bf APPENDIX A: Complete self-consistency equation}\\\\ \indent If we systematically
integrate out all the ligand degrees of freedom, we obtain the following
self-consistency condition which constitutes the lattice nature of the
problem. The physical meaning of each term can be diagramatically
illustrated in t he same manner as we showed in sec.III: \begin{eqnarray*}
{\cal
G}_{AA}^{\uparrow\uparrow}(i\omega_{n})&=&i\omega_{n}-\epsilon_{d}+\mu+h_{z}-\frac{U}{2}-t^{2}G_{BB}^{\uparrow\uparrow}(i\omega_{n})\\
&-& t^{2}\xi_{AB}^{\uparrow\downarrow}\xi_{BA}^{\downarrow\uparrow}G_{BB}^{\downarrow\downarrow}(i\omega_{n})\\
&+& t^{2} (\xi_{AB}^{\uparrow\downarrow}F_{BB}^{\downarrow\uparrow}(i\omega_{n})+\xi_{BA}^{\downarrow\uparrow}F_{BB}^{\uparrow\downarrow}(i\omega_{n}))\\
{\cal
G}_{BB}^{\uparrow\uparrow}(i\omega_{n})&=&i\omega_{n}-\epsilon_{d}+\mu-h_{z}-\frac{U}{2}-t^{2}G_{AA}^{\uparrow\uparrow}(i\omega_{n})\\
&-& t^{2}\xi_{BA}^{\uparrow\downarrow}\xi_{AB}^{\downarrow\uparrow}G_{AA}^{\downarrow\downarrow}(i\omega_{n})\\
&+& t^{2} (\xi_{BA}^{\uparrow\downarrow}F_{AA}^{\downarrow\uparrow}(i\omega_{n})+\xi_{AB}^{\downarrow\uparrow}F_{AA}^{\uparrow\downarrow}(i\omega_{n}))\\
{\cal
G}_{AA}^{\downarrow\downarrow}(i\omega_{n})&=&i\omega_{n}+\epsilon_{d}-\mu+h_{z}+\frac{U}{2}-t^{2}G_{BB}^{\downarrow\downarrow}(i\omega_{n})\\
&-& t^{2}\xi_{AB}^{\downarrow\uparrow}\xi_{BA}^{\uparrow\downarrow}G_{BB}^{\uparrow\uparrow}(i\omega_{n})\\
&-& t^{2} (\xi_{BA}^{\uparrow\downarrow}F_{BB}^{\downarrow\uparrow}(i\omega_{n})+\xi_{AB}^{\downarrow\uparrow}F_{BB}^{\uparrow\downarrow}(i\omega_{n}))\\
{\cal
G}_{BB}^{\downarrow\downarrow}(i\omega_{n})&=&i\omega_{n}+\epsilon_{d}-\mu-h_{z}+\frac{U}{2}-t^{2}G_{AA}^{\downarrow\downarrow}(i\omega_{n})\\
&-& t^{2}\xi_{BA}^{\downarrow\uparrow}\xi_{AB}^{\uparrow\downarrow}G_{AA}^{\uparrow\uparrow}(i\omega_{n})\\
&-& t^{2} (\xi_{AB}^{\uparrow\downarrow}F_{AA}^{\downarrow\uparrow}(i\omega_{n})+\xi_{BA}^{\downarrow\uparrow}F_{AA}^{\uparrow\downarrow}(i\omega_{n}))\\\\\\
{\cal
G}_{AB}^{\uparrow\uparrow}(i\omega_{n})&=&t^{\prime}-t^{2}G_{BA}^{\uparrow\uparrow}(i\omega_{n})\\
&-& t^{2}\xi_{AB}^{\uparrow\downarrow}\xi_{AB}^{\downarrow\uparrow}G_{BA}^{\downarrow\downarrow}(i\omega_{n})\\
&+& t^{2} (\xi_{AB}^{\uparrow\downarrow}F_{BA}^{\downarrow\uparrow}(i\omega_{n})+\xi_{AB}^{\downarrow\uparrow}F_{BA}^{\uparrow\downarrow}(i\omega_{n}))\\
{\cal
G}_{BA}^{\uparrow\uparrow}(i\omega_{n})&=&t^{\prime}-t^{2}G_{AB}^{\uparrow\uparrow}(i\omega_{n})\\
&-& t^{2}\xi_{BA}^{\downarrow\uparrow}\xi_{BA}^{\uparrow\downarrow}G_{AB}^{\downarrow\downarrow}(i\omega_{n})\\
&+& t^{2} (\xi_{BA}^{\downarrow\uparrow}F_{AB}^{\uparrow\downarrow}(i\omega_{n})+\xi_{BA}^{\uparrow\downarrow}F_{AB}^{\downarrow\uparrow}(i\omega_{n}))\\
{\cal
G}_{AB}^{\downarrow\downarrow}(i\omega_{n})&=&t^{\prime}-t^{2}G_{BA}^{\downarrow\downarrow}(i\omega_{n})\\
&-& t^{2}\xi_{AB}^{\uparrow\downarrow}\xi_{AB}^{\downarrow\uparrow}G_{BA}^{\uparrow\uparrow}(i\omega_{n})\\
&-&t^{2} (\xi_{AB}^{\uparrow\downarrow}F_{BA}^{\downarrow\uparrow}(i\omega_{n})+\xi_{AB}^{\downarrow\uparrow}F_{BA}^{\uparrow\downarrow}(i\omega_{n}))\\
{\cal
G}_{BA}^{\downarrow\downarrow}(i\omega_{n})&=&t^{\prime}-t^{2}G_{AB}^{\downarrow\downarrow}(i\omega_{n})\\
&-& t^{2}\xi_{BA}^{\downarrow\uparrow}\xi_{BA}^{\uparrow\downarrow}G_{AB}^{\uparrow\uparrow}(i\omega_{n})\\
&-&t^{2} (\xi_{BA}^{\downarrow\uparrow}F_{AB}^{\uparrow\downarrow}(i\omega_{n})+\xi_{BA}^{\uparrow\downarrow}F_{AB}^{\downarrow\uparrow}(i\omega_{n}))\\\\\\
{\cal
F}_{AA}^{\uparrow\downarrow}(i\omega_{n})&=&t^{2}F_{BB}^{\uparrow\downarrow}(i\omega_{n})\\
&-& t^{2}\xi_{AB}^{\uparrow\downarrow}\xi_{BA}^{\uparrow\downarrow}F_{BB}^{\downarrow\uparrow}(i\omega_{n})\\
&-&t^{2} (\xi_{AB}^{\uparrow\downarrow}G_{BB}^{\downarrow\downarrow}(i\omega_{n})-\xi_{BA}^{\uparrow\downarrow}G_{BB}^{\uparrow\uparrow}(i\omega_{n}))\\
{\cal
F}_{AA}^{\downarrow\uparrow}(i\omega_{n})&=&t^{2}F_{BB}^{\downarrow\uparrow}(i\omega_{n})\\
&-&t^{2}\xi_{BA}^{\downarrow\uparrow}\xi_{AB}^{\downarrow\uparrow}F_{BB}^{\uparrow\downarrow}(i\omega_{n})\\
&-& t^{2} (\xi_{BA}^{\downarrow\uparrow}G_{BB}^{\downarrow\downarrow}(i\omega_{n})-\xi_{AB}^{\downarrow\uparrow}G_{BB}^{\uparrow\uparrow}(i\omega_{n}))\\
{\cal
F}_{BB}^{\uparrow\downarrow}(i\omega_{n})&=&t^{2}F_{AA}^{\uparrow\downarrow}(i\omega_{n})\\
&-& t^{2}\xi_{AB}^{\uparrow\downarrow}\xi_{BA}^{\uparrow\downarrow}F_{AA}^{\downarrow\uparrow}(i\omega_{n})\\
&-&t^{2} (\xi_{BA}^{\uparrow\downarrow}G_{AA}^{\downarrow\downarrow}(i\omega_{n})-\xi_{AB}^{\uparrow\downarrow}G_{AA}^{\uparrow\uparrow}(i\omega_{n}))\\
{\cal
F}_{BB}^{\downarrow\uparrow}(i\omega_{n})&=&t^{2}F_{AA}^{\downarrow\uparrow}(i\omega_{n})\\
&-& t^{2}\xi_{AB}^{\downarrow\uparrow}\xi_{BA}^{\downarrow\uparrow}F_{AA}^{\uparrow\downarrow}(i\omega_{n})\\
&-&t^{2} (\xi_{AB}^{\downarrow\uparrow}G_{AA}^{\downarrow\downarrow}(i\omega_{n})-\xi_{BA}^{\downarrow\uparrow}G_{AA}^{\uparrow\uparrow}(i\omega_{n}))\\\\\\
{\cal
F}_{AB}^{\uparrow\downarrow}(i\omega_{n})&=&-\eta+t^{2}F_{BA}^{\uparrow\downarrow}(i\omega_{n})\\
&-& t^{2} (\xi_{AB}^{\uparrow\downarrow})^{2}F_{BA}^{\downarrow\uparrow}(i\omega_{n})\\
&-& t^{2}\xi_{AB}^{\uparrow\downarrow}(G_{BA}^{\downarrow\downarrow}(i\omega_{n})-G_{BA}^{\uparrow\uparrow}(i\omega_{n}))\\
{\cal
F}_{BA}^{\downarrow\uparrow}(i\omega_{n})&=&-\eta+t^{2}F_{AB}^{\downarrow\uparrow}(i\omega_{n})\\
&-& t^{2} (\xi_{BA}^{\downarrow\uparrow})^{2}F_{AB}^{\uparrow\downarrow}(i\omega_{n})\\
&-& t^{2}\xi_{BA}^{\downarrow\uparrow}(G_{AB}^{\downarrow\downarrow}(i\omega_{n})-G_{AB}^{\uparrow\uparrow}(i\omega_{n}))\\
{\cal
F}_{BA}^{\uparrow\downarrow}(i\omega_{n})&=&-\eta+t^{2}F_{AB}^{\uparrow\downarrow}(i\omega_{n})\\
&-& t^{2} (\xi_{BA}^{\uparrow\downarrow})^{2}F_{AB}^{\downarrow\uparrow}(i\omega_{n})\\
&-& t^{2}\xi_{BA}^{\uparrow\downarrow}(G_{AB}^{\downarrow\downarrow}(i\omega_{n})-G_{AB}^{\uparrow\uparrow}(i\omega_{n}))\\
{\cal
F}_{AB}^{\downarrow\uparrow}(i\omega_{n})&=&-\eta+t^{2}F_{BA}^{\downarrow\uparrow}(i\omega_{n})\\
&-& t^{2} (\xi_{AB}^{\downarrow\uparrow})^{2}F_{BA}^{\uparrow\downarrow}(i\omega_{n})\\
&-& t^{2}\xi_{AB}^{\downarrow\uparrow}(G_{BA}^{\downarrow\downarrow}(i\omega_{n})-G_{BA}^{\uparrow\uparrow}(i\omega_{n}))
\end{eqnarray*}\\\\
 {\bf APPENDIX B: Manifestation of a
sublattice symmetry in the formula of the accept ratio}\\\\ \indent Since
the Nambu-Gorkov representation is used in (9), we first need to transform
to the usual representation.  This can be done by transforming only the
down spin components in the following way:\\ \begin{eqnarray*}
G_{AA}^{\downarrow\downarrow}(\tau_{i},\tau_{i}) \longmapsto
1-G_{AA}^{\downarrow\downarrow}(\tau_{i},\tau_{i}) \nonumber\\
G_{BB}^{\downarrow\downarrow}(\tau_{i},\tau_{i}) \longmapsto
1-G_{BB}^{\downarrow\downarrow}(\tau_{i},\tau_{i}) \nonumber\\
\end{eqnarray*} This will transform the accept ratio to the following
form:\\ \begin{eqnarray*}
R_{A}(\tau_{i})=((e^{\lambda\sigma_{A}(\tau_{i})}-e^{-\lambda\sigma_{A}(\tau_{i})})G_{AA}^{\uparrow\uparrow}(\tau_{i},\tau_{i})+e^{-\lambda\sigma_{A}(\tau_{i})})\\
\times((e^{-\lambda\sigma_{A}(\tau_{i})}-e^{\lambda\sigma_{A}(\tau_{i})})G_{AA}^{\downarrow\downarrow}(\tau_{i},\tau_{i})+e^{\lambda\sigma_{A}(\tau_{i})})\\
-(e^{\lambda\sigma_{A}(\tau_{i})}-e^{-\lambda\sigma_{A}(\tau_{i})})^{2}F_{AA}^{\uparrow\downarrow}(\tau_{i},\tau_{i})F_{AA}^{\downarrow\uparrow}(\tau_{i},\tau_{i})\\\\
R_{B}(\tau_{i})=((e^{\lambda\sigma_{B}(\tau_{i})}-e^{-\lambda\sigma_{B}(\tau_{i})})G_{BB}^{\uparrow\uparrow}(\tau_{i},\tau_{i})+e^{-\lambda\sigma_{B}(\tau_{i})})\\
\times((e^{-\lambda\sigma_{B}(\tau_{i})}-e^{\lambda\sigma_{B}(\tau_{i})})G_{BB}^{\downarrow\downarrow}(\tau_{i},\tau_{i})+e^{\lambda\sigma_{B}(\tau_{i})})\\
-(e^{\lambda\sigma_{B}(\tau_{i})}-e^{-\lambda\sigma_{B}(\tau_{i})})^{2}F_{BB}^{\uparrow\downarrow}(\tau_{i},\tau_{i})F_{BB}^{\downarrow\uparrow}(\tau_{i},\tau_{i}) 
\end{eqnarray*} in which the sublattice symmetry is manifestly
apparent.\\\\
 {\bf APPENDIX C: Calculation of
$\chi_{\vec{Q}}^{z}(\omega)$}\\\\ In this appendix, we show the details of
the procedure to compute a staggered spin susceptibility
$\chi_{\vec{Q}}^{z}(\omega)$ from an imaginary-time causal staggered
spin-spin correlation function which is given as:  \begin{eqnarray*}
S_{z}(\vec{Q},\tau-\tau^{\prime})&=&<\hat{M}_{z}(\tau)\hat{M}_{z}(\tau^{\prime})>\\
&=&\frac{\int_{local}D\hat{\psi}D\hat{\psi}^{\dagger}\hat{M}_{z}(\tau)\hat{M}_{z}(\tau^{\prime})e^{-S_{eff}}}{Z_{MF}}
\end{eqnarray*} The above trace involves four point correlators and a Wick
contraction needs to be performed. Since the effective action is
bilinearized by the Hubbard-Stratonovich transformation, the functional
integral over the fermionic degrees of freedom can be exactly performed
and the above trace simply becomes a sum of the products of local Green's
functions averaged over the Ising variable. In other words, all the vertex
corrections at the local level are decoupled by means of
Hubbard-Stratonovich transformation and re-absorbed into the dependencies
of the local propagators on the Ising variable. For example, the term
$<\hat{n}_{A\downarrow}(\tau)\hat{n}_{B\downarrow}(\tau)>$ can be
integrated in the following way:\\ \begin{eqnarray*}
<\hat{n}_{A\downarrow}(\tau)\hat{n}_{B\downarrow}(\tau^{\prime})>&=&<\hat{d}_{A\downarrow}^{\dagger}(\tau)\hat{d}_{A\downarrow}(\tau)\hat{d}_{B\downarrow}^{\dagger}(\tau^{\prime})\hat{d}_{B\downarrow}(\tau^{\prime}
)>\\
&=&<G_{AA}^{\downarrow\downarrow\sigma(l)}(\tau,\tau)G_{BB}^{\downarrow\downarrow\sigma(l)}(\tau^{\prime},\tau^{\prime})>_{\sigma(l)}\\
&
&-<G_{AB}^{\downarrow\downarrow\sigma(l)}(\tau,\tau^{\prime})G_{BA}^{\downarrow\downarrow\sigma(l)}(\tau^{\prime},\tau)>_{\sigma(l)}
\end{eqnarray*} where $< ... >_{\sigma(l)}$ means an averaging over the
Ising variable. The normal order for down spin components is reversed due
to the Nambu representation. The full expression of $S_{z}(\vec{Q},\tau)$
becomes quite lengthy and is omitted here \cite{watanabe2}. The averaging
over the Ising variable is performed by the QMC sampling. Since
$S_{z}(\vec{Q},\tau)$ is a scattering function(i.e., fluctuation of a spin
degrees of freedom), the transfer function which defines the relation
between the imaginary-time and real-time quantities becomes:\\
\begin{eqnarray*}
S_{z}(\vec{Q},\tau)=\frac{1}{2\pi}\int_{-\infty}^{+\infty}d\omega
e^{-\tau\omega}S_{z}(\vec{Q},\omega)  \end{eqnarray*} where
$S_{z}(\vec{Q},\omega)$ must satisfy the equation of detailed balance
$S_{z}(\vec{Q},-\omega) = e^{-\beta\omega}S_{z}(\vec{Q},\omega)$. (Please
note the difference from the fermion operators (10).) Here, the equation
of detailed balance can be used as a constraint for the default model in
Maximum Entropy \cite{deisz1}
 or simply absorb it into the transfer
function, which is the approach we adopted here.\\\\\\\\

\pagebreak

{\bf Figure 1 :} A schematic picture of our model.  Elliptic curves define
the two-site cluster.  All sites have the same nonzero $U$, all
inter-cluster hopping have the same $t$, and all intra-cluster hopping
have the same $t^{\prime}$.

{\bf Figure 2 :} Evolution of single-particle spectra as a function of
temperature without inter-cluster anomalous terms. (a) is for a small
doping($\delta \sim 0.05$) and (b) for a large doping($\delta \sim 0.17$). 

{\bf Figure 3 :} Single-particle spectrum in the normal phase for
$\beta=16$ and $\delta \sim 0.05$. 

{\bf Figure 4 :} Evolution of single-particle spectra as a function of
temperature with inter-cluster anomalous terms. (a) is
for  small doping($\delta \sim 0.04$) and (b) for  large doping($\delta
\sim 0.17$). 

{\bf Figure 5 :} DOS of normal(solid line) and SC(dashed line) state for
$\mu = 0.7$ as a function of temperature.  A gap feature clearly builds up
as the \SCg correlation develops. 

{\bf Figure 6 :} A single-particle gap(diamond), Cooper pair
density($\propto |F(0)|^{2}$, cross) and $T_{c}$(square) vs. doping at
$\beta = 16$. Vertical units are arbitrary.

{\bf Figure 7 :} $\chi_{z}(\vec{Q},\omega)$ as a function of temperature
for the underdoped region. (a):without inter-cluster pairing and (b):with
inter-cluster pairing. 

{\bf Figure 8 :} $\chi_{z}(\vec{Q},\omega)$ as a function of temperature
near the optimal doping region. (a):without inter-cluster pairing and (b):with
inter-cluster pairing. 

{\bf Figure 9 :} $\chi_{z}(\vec{Q},\omega)$ as a function of doping.
(a):without inter-cluster pairing and (b):with inter-cluster pairing. 

{\bf Figure 10 :} $\chi_{z}(\vec{Q},\omega)$ spectra of normal(dash line)
and SC(solid line) for $\beta = 16$ and $\mu = 0.3$. Coherence is
severely degraded and the gap is filled in the normal phase. 

{\bf Figure 11 :} $Re[\sigma^{para}(\omega)]$ as a function of temperature
for $\mu = 0.7$(a) and $\mu = 0.3$(b). The horizontal unit is $2t$ and the
vertical unit is arbitrary. 

{\bf Figure 12 :} $\chi_{d}(\omega)$ as a function of temperature for $\mu
= 0.7$(a) and $\mu = 0.3$(b).  The horizontal unit is $2t$ and the vertical
unit is arbitrary. 

{\bf Figure 13 :} (a)$M_{z}$ vs. temperature at half filling. Units are
$\mu_{B}$ and {2t} for vertical and horizontal axis, respectively.
(b)$|F(0)|^{2}$ vs. doping at $\beta = 16$. A unit for the vertical axis
is arbitrary. $|F(0)|^{2}$ vs. temperature for $\mu = 0.7$(c)  and $\mu =
0.3$(d). 

{\bf Figure 14 :} Schematic phase diagram obtained based on our model.
The onset of SC was determined by the appearance of a
thermal average of an instantaneous pairing amplitude. Error bars
correspond to the size of increments of sampling points we studied.

\begin{figure}
\includegraphics{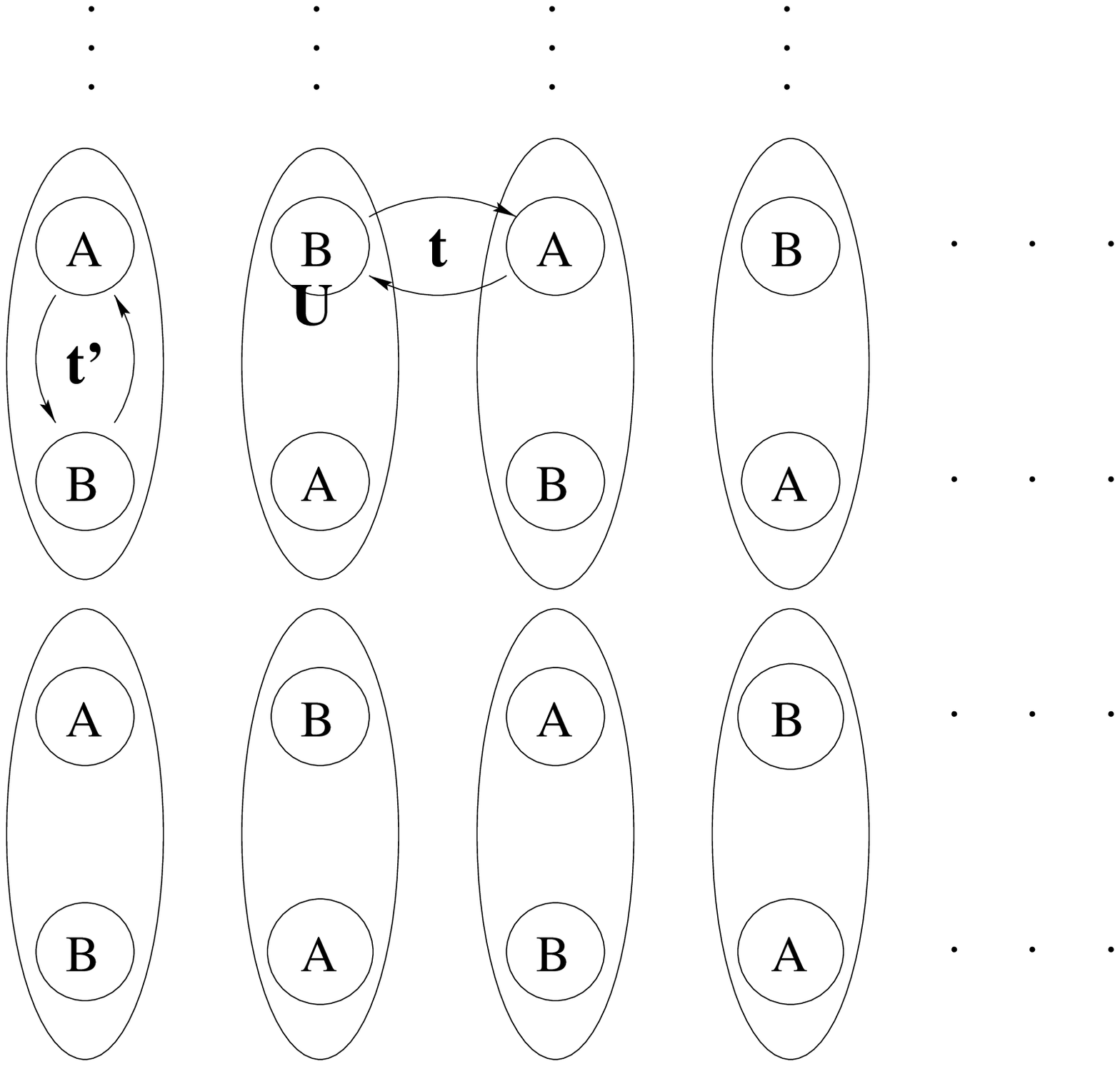}
{\bf Figure 1 :} A schematic picture of our model.  Elliptic curves define
the two-site cluster.  All sites have the same nonzero $U$, all
inter-cluster hopping have the same $t$, and all intra-cluster hopping
have the same $t^{\prime}$. 
\end{figure}

\begin{figure}
\includegraphics{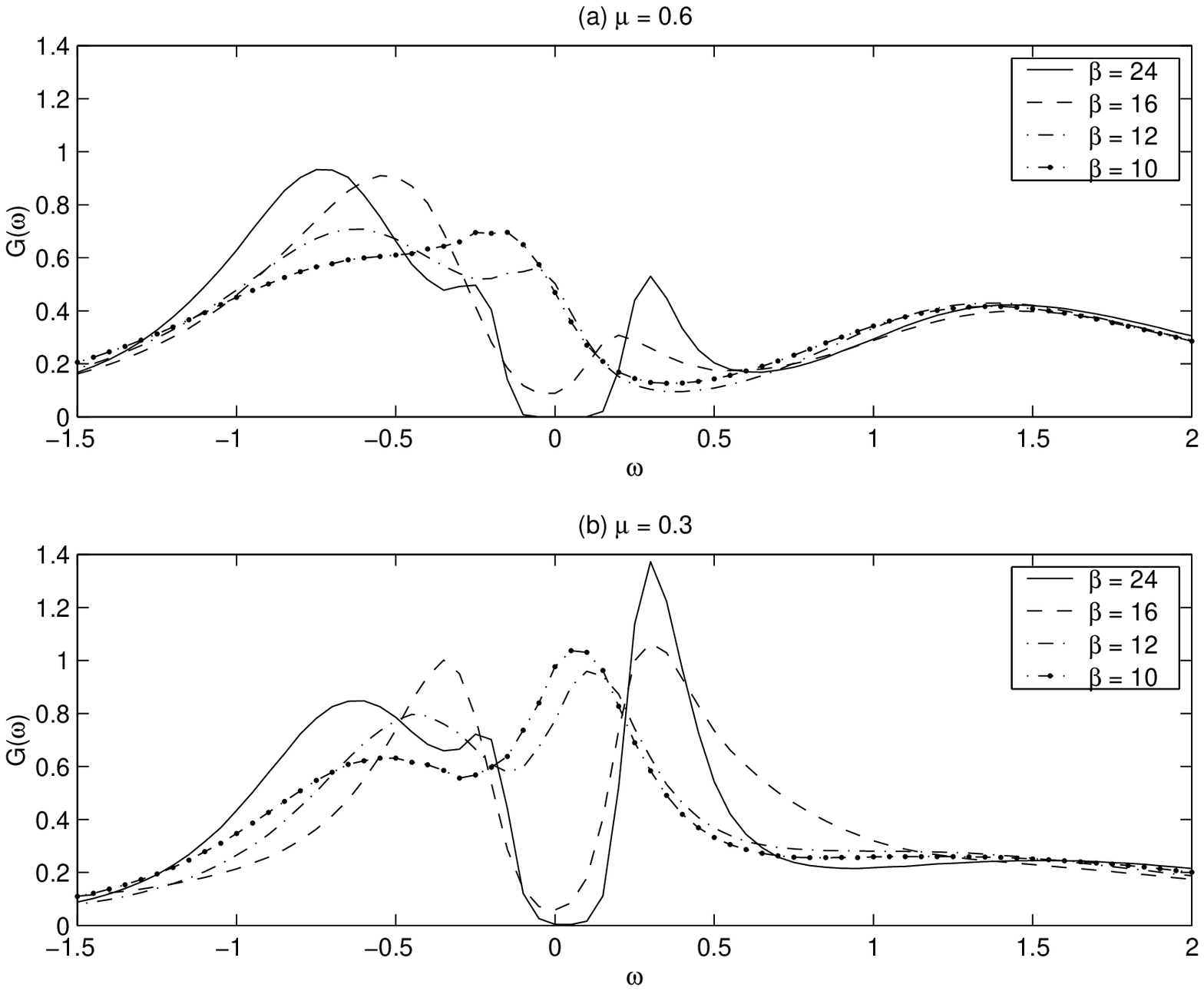}
{\bf Figure 2 :} Evolution of single-particle spectra as a function of
temperature without inter-cluster anomalous terms. (a) is for a small
doping($\delta \sim 0.05$) and (b) for a large doping($\delta \sim 0.17$). 
\end{figure}

\begin{figure}
\includegraphics{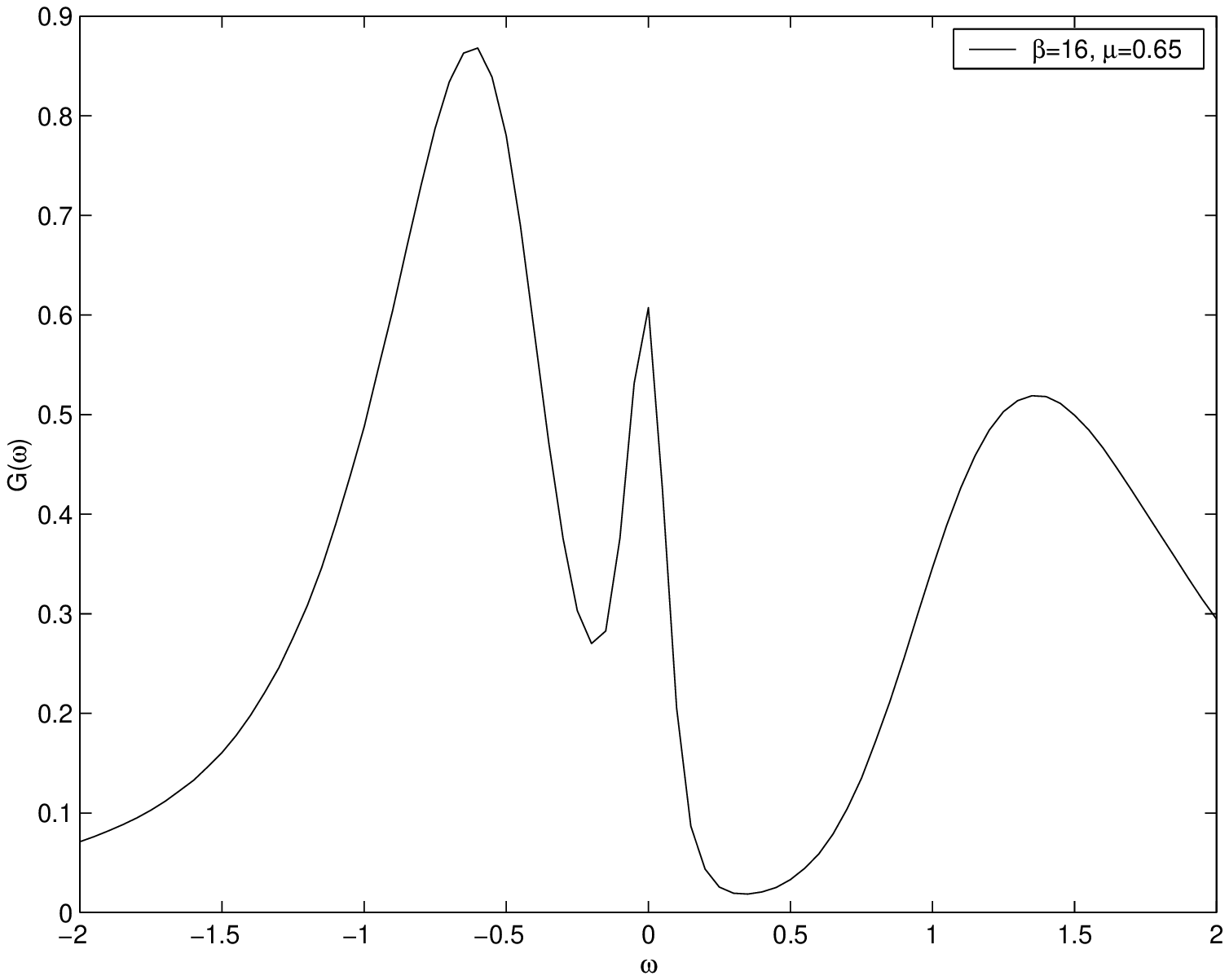}

{\bf Figure 3 :} Single-particle spectrum in the normal phase for
$\beta=16$ and $\delta \sim 0.05$. 
\end{figure}

\begin{figure}
\includegraphics{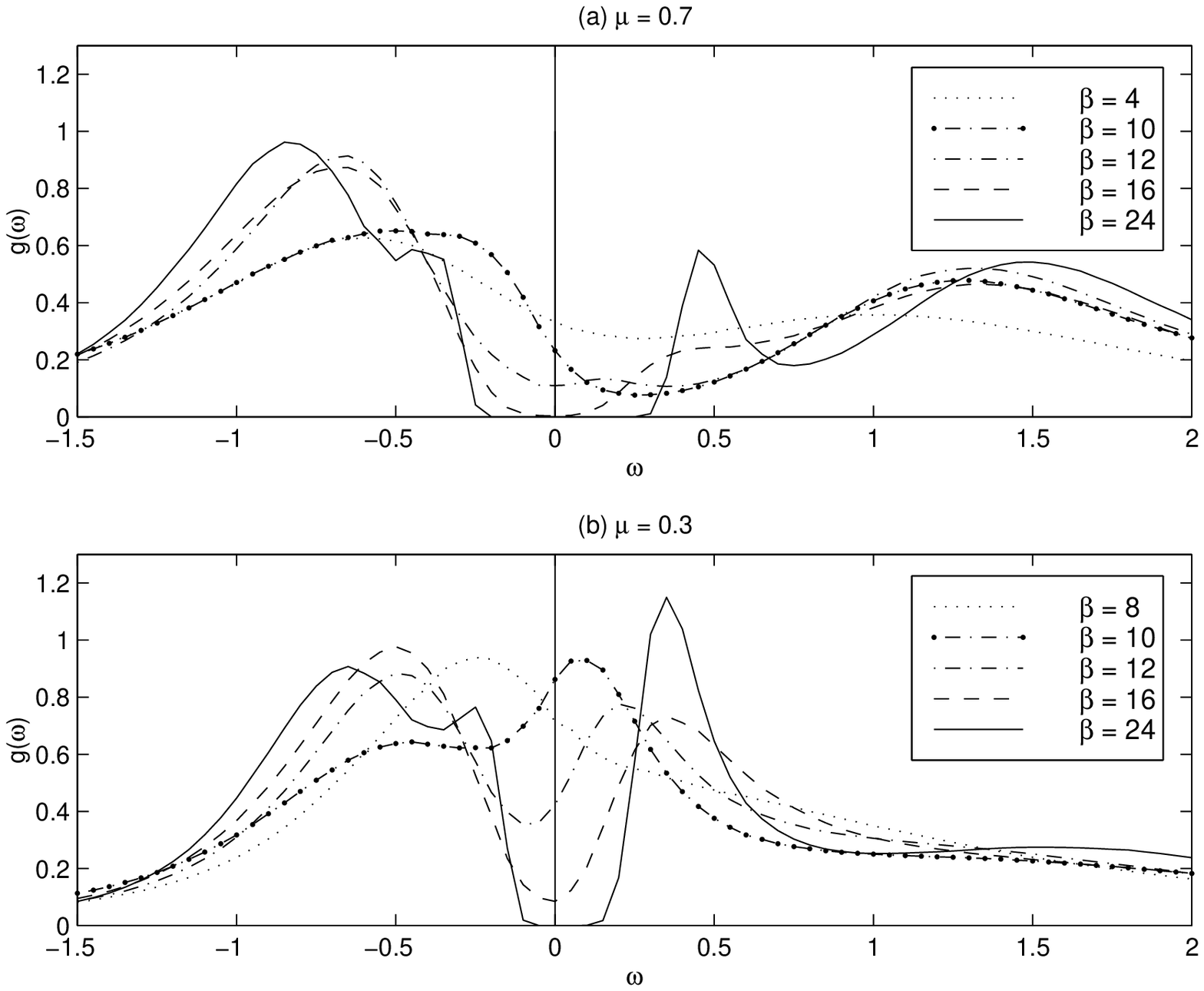}
{\bf Figure 4 :} Evolution of single-particle spectra as a function of
temperature with inter-cluster anomalous terms. (a) is
for small doping($\delta \sim 0.04$) and (b) for large doping($\delta
\sim 0.17$). 
\end{figure}

\begin{figure}
\includegraphics{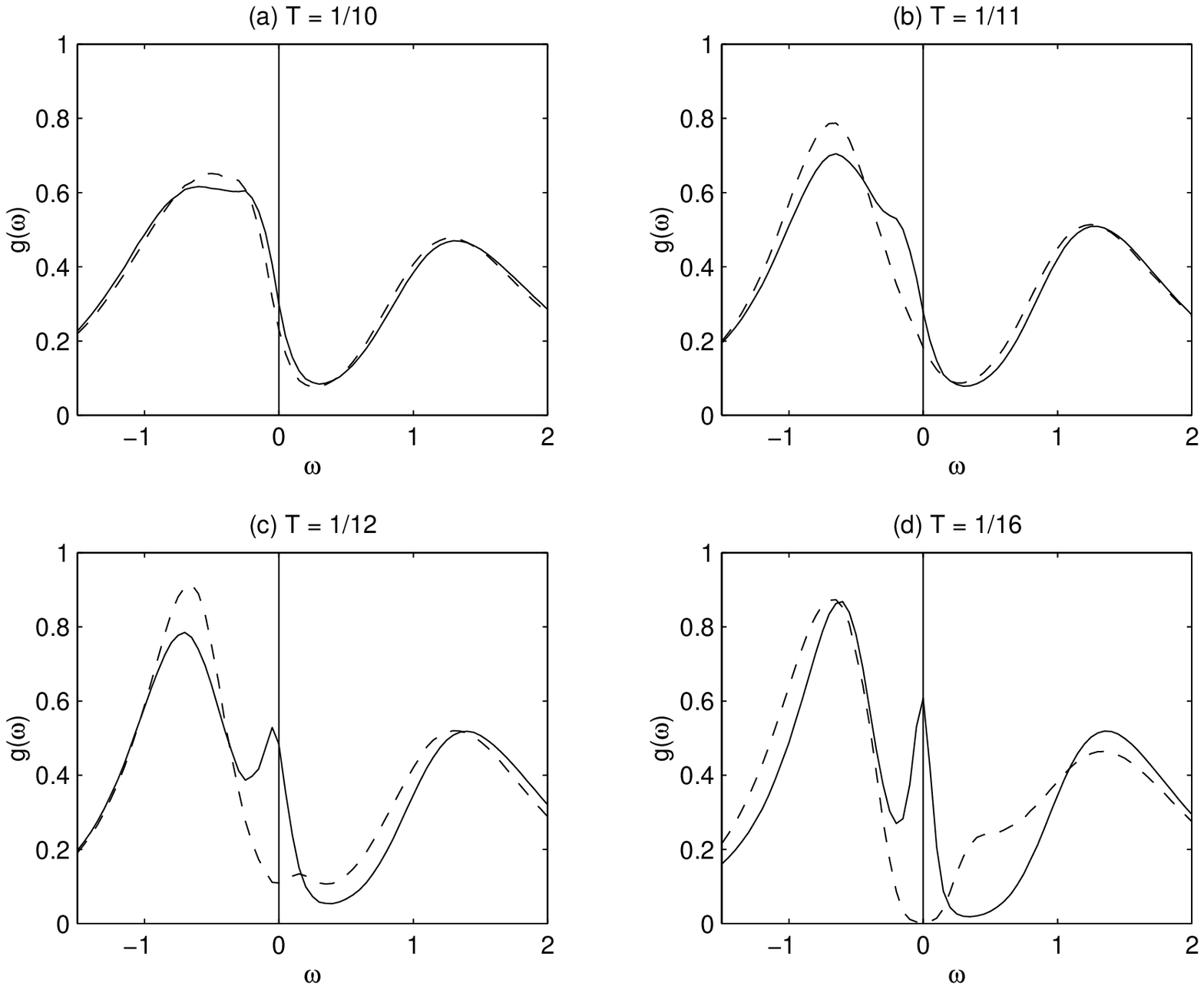}
{\bf Figure 5 :} DOS of normal(solid line) and SC(dashed line) state for
$\mu = 0.7$ as a function of temperature.  A gap feature clearly builds up
as the \SCg correlation develops. 
\end{figure}

\begin{figure}
\includegraphics{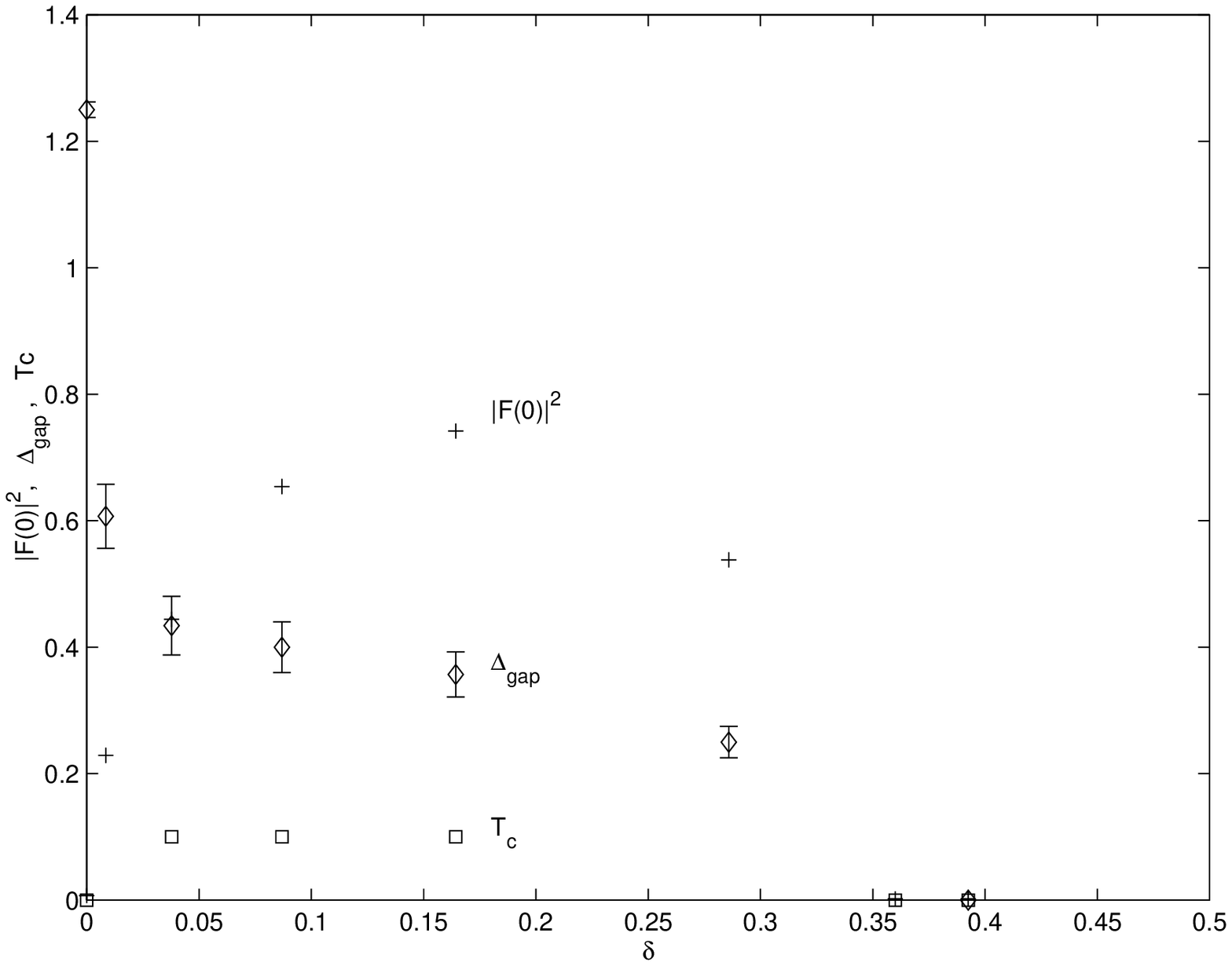}
{\bf Figure 6 :} A single-particle gap(diamond), Cooper pair
density($\propto |F(0)|^{2}$, cross) and $T_{c}$(square) vs. doping at
$\beta = 16$. Vertical units are arbitrary.
\end{figure}

\begin{figure}
\includegraphics{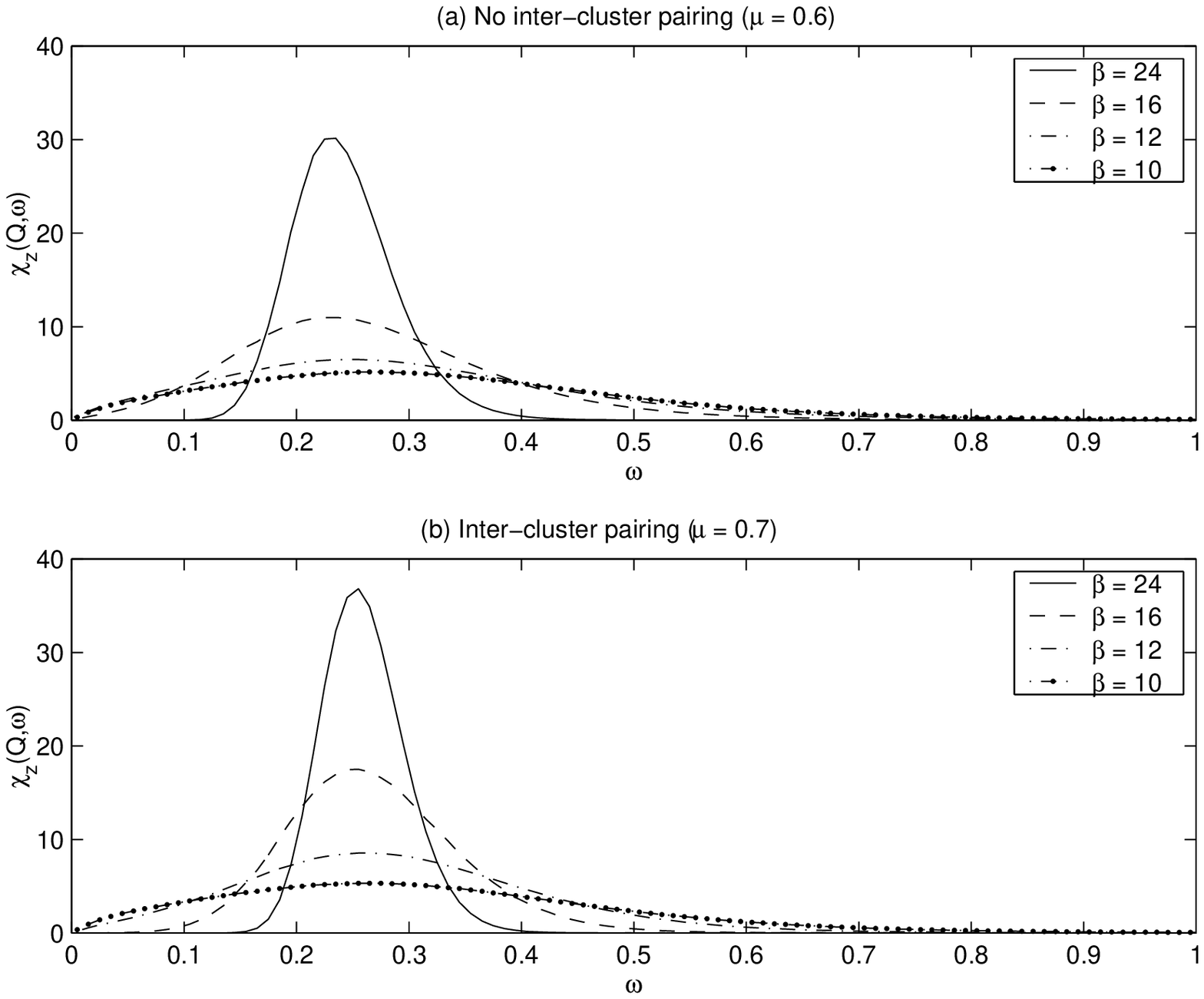}
{\bf Figure 7 :} $\chi_{z}(\vec{Q},\omega)$ as a function of temperature
for the underdoped region. (a):without inter-cluster pairing and (b):with
inter-cluster pairing. 
\end{figure}

\begin{figure}
\includegraphics{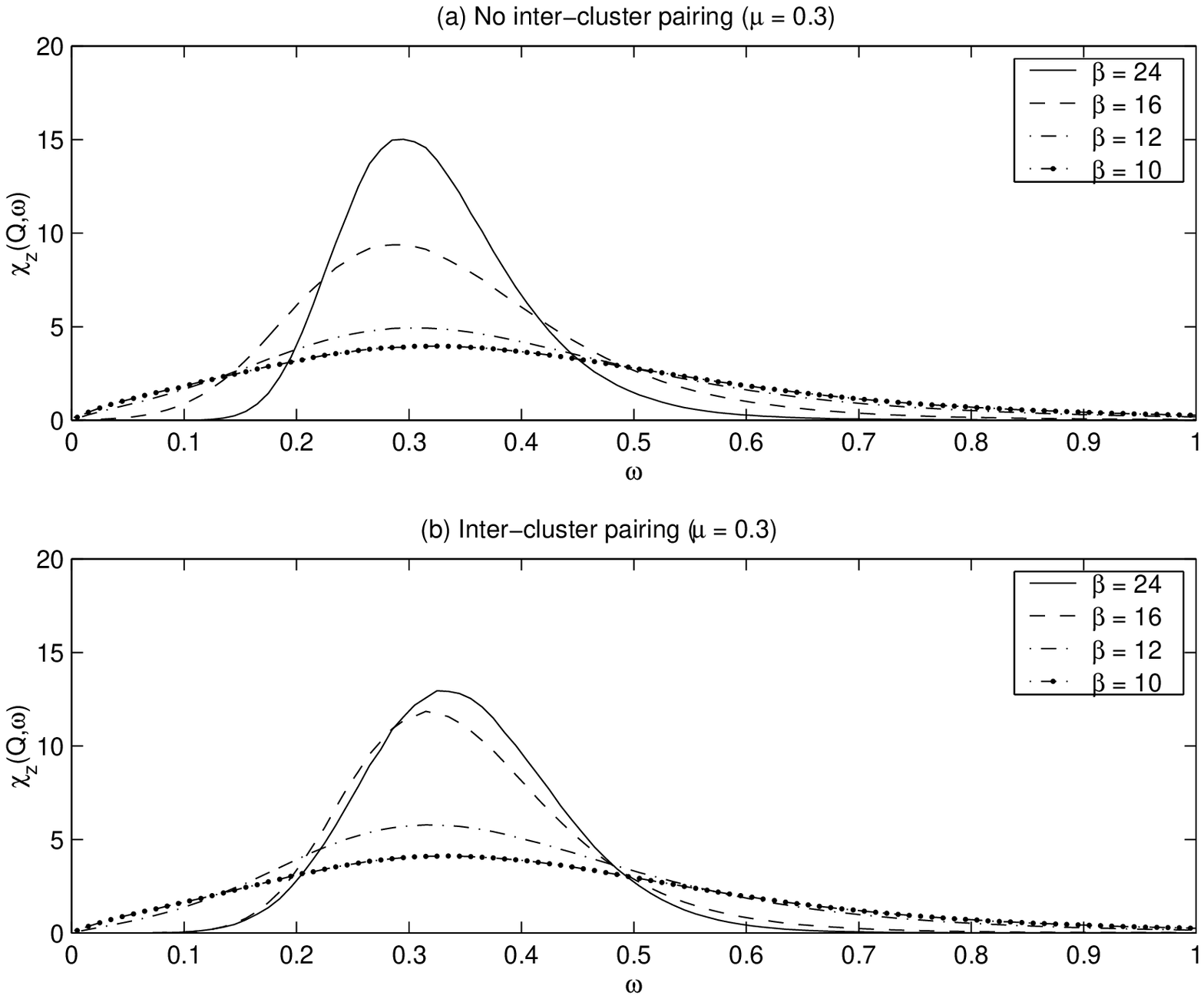}
{\bf Figure 8 :} $\chi_{z}(\vec{Q},\omega)$ as a function of temperature
near optimal doping region. (a):without inter-cluster pairing and (b):with
inter-cluster pairing. 
\end{figure}

\begin{figure}
\includegraphics{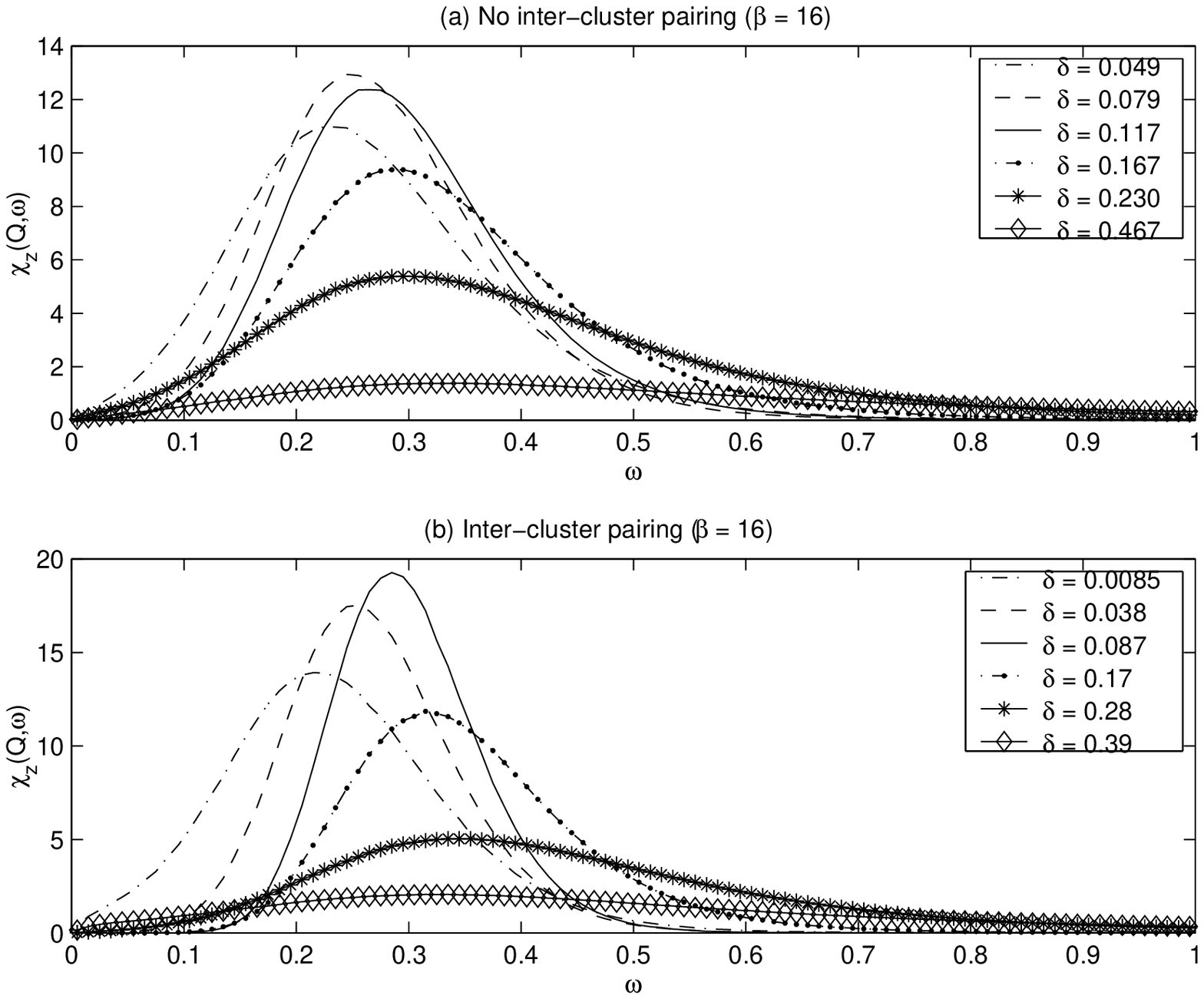}
{\bf Figure 9 :} $\chi_{z}(\vec{Q},\omega)$ as a function of doping.
(a):without inter-cluster pairing and (b):with inter-cluster pairing. 
\end{figure}

\begin{figure}
\includegraphics{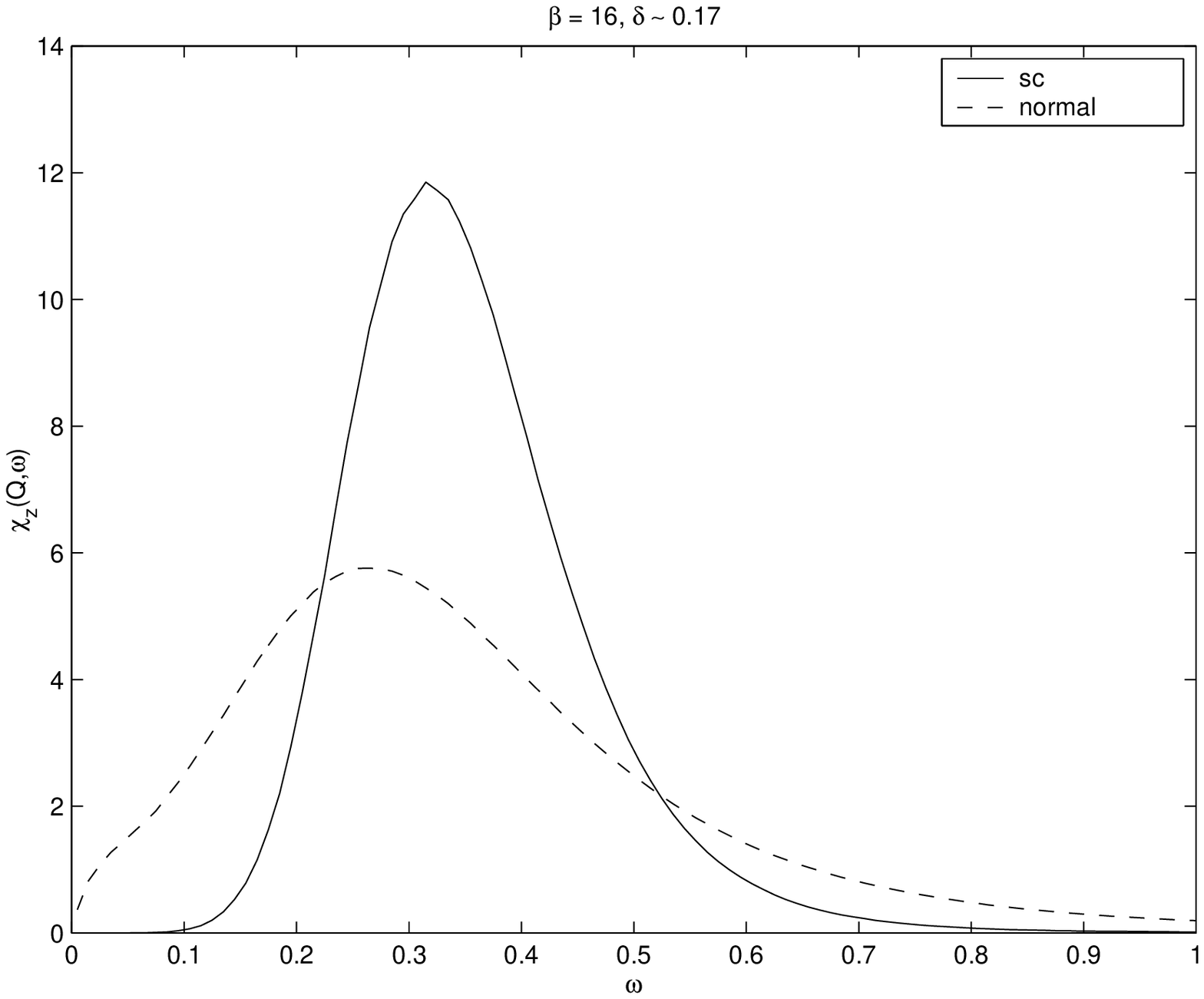}
{\bf Figure 10 :} $\chi_{z}(\vec{Q},\omega)$ spectra of normal(dash line)
and SC(solid line) for $\beta = 16$ and $\mu = 0.3$. Coherence is
severely degraded and the gap is filled in the normal phase. 
\end{figure}

\begin{figure}
\includegraphics{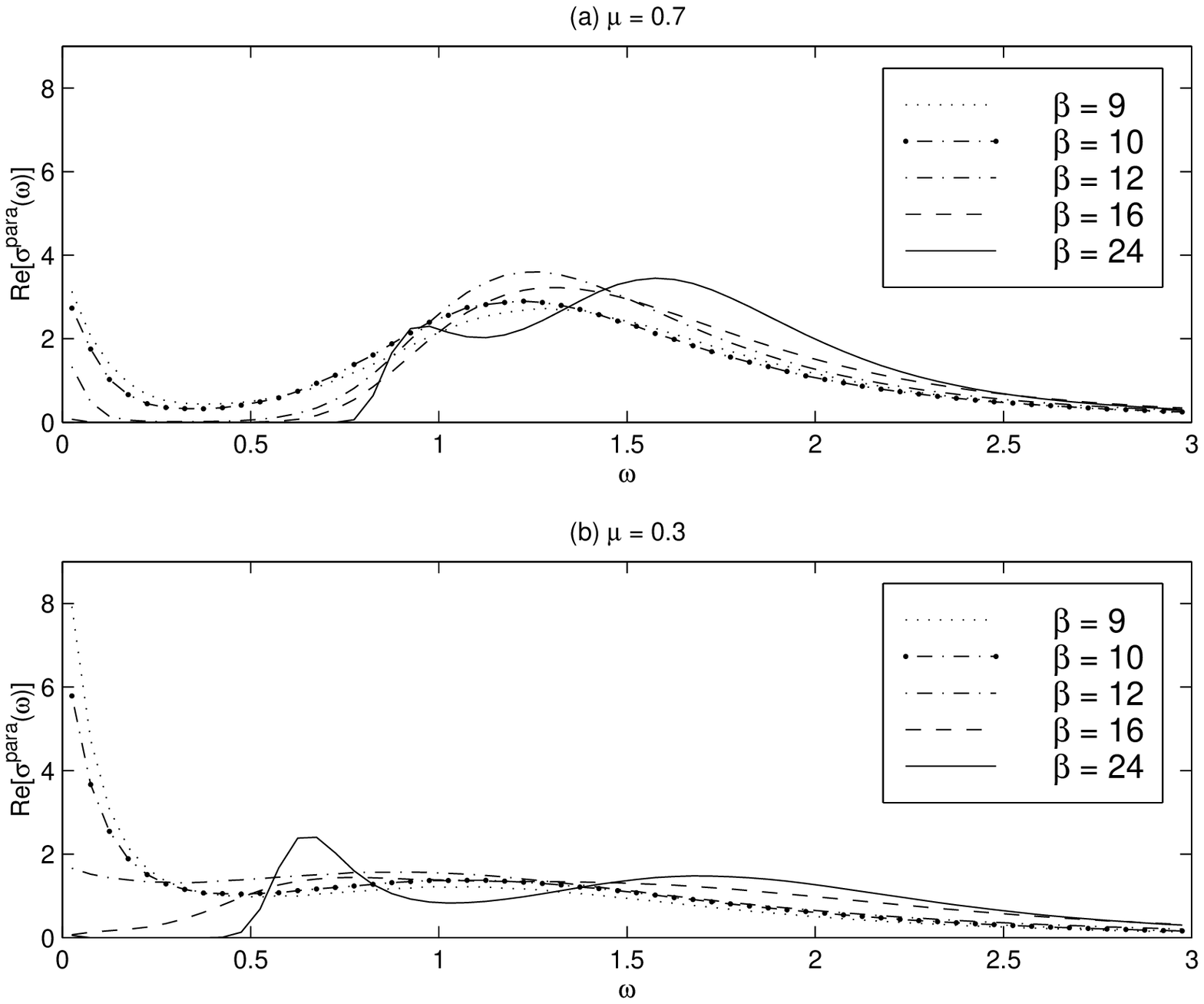}
{\bf Figure 11 :} $Re[\sigma^{para}(\omega)]$ as a function of temperature
for $\mu = 0.7$(a) and $\mu = 0.3$(b). The horizontal unit is $2t$ and the
vertical unit is arbitrary. 
\end{figure}

\begin{figure}
\includegraphics{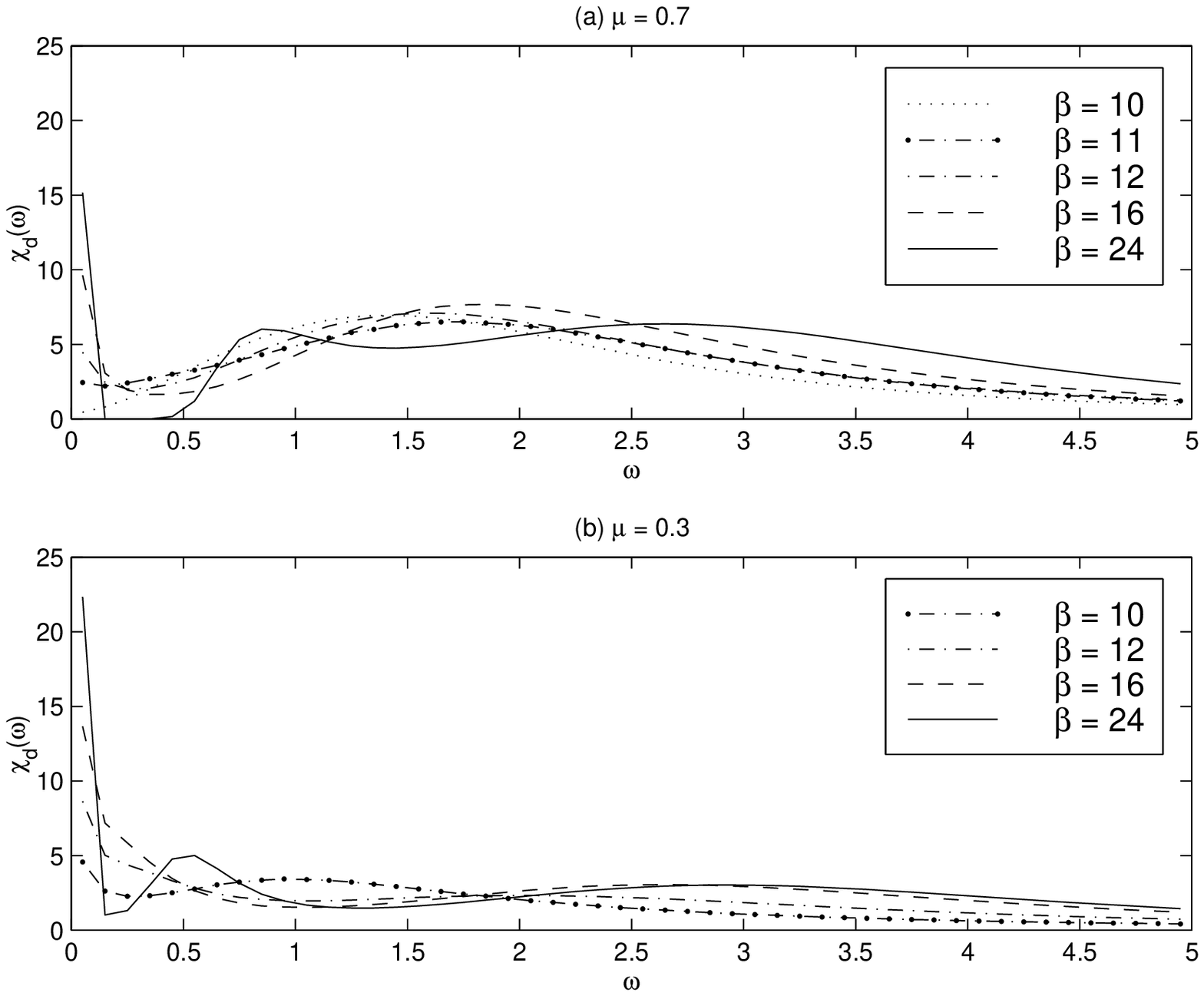}
{\bf Figure 12 :} $\chi_{d}(\omega)$ as a function of temperature for $\mu
= 0.7$(a) and $\mu = 0.3$(b).  The horizontal unit is $2t$ and the
vertical unit is arbitrary. 
\end{figure}

\begin{figure}
\includegraphics{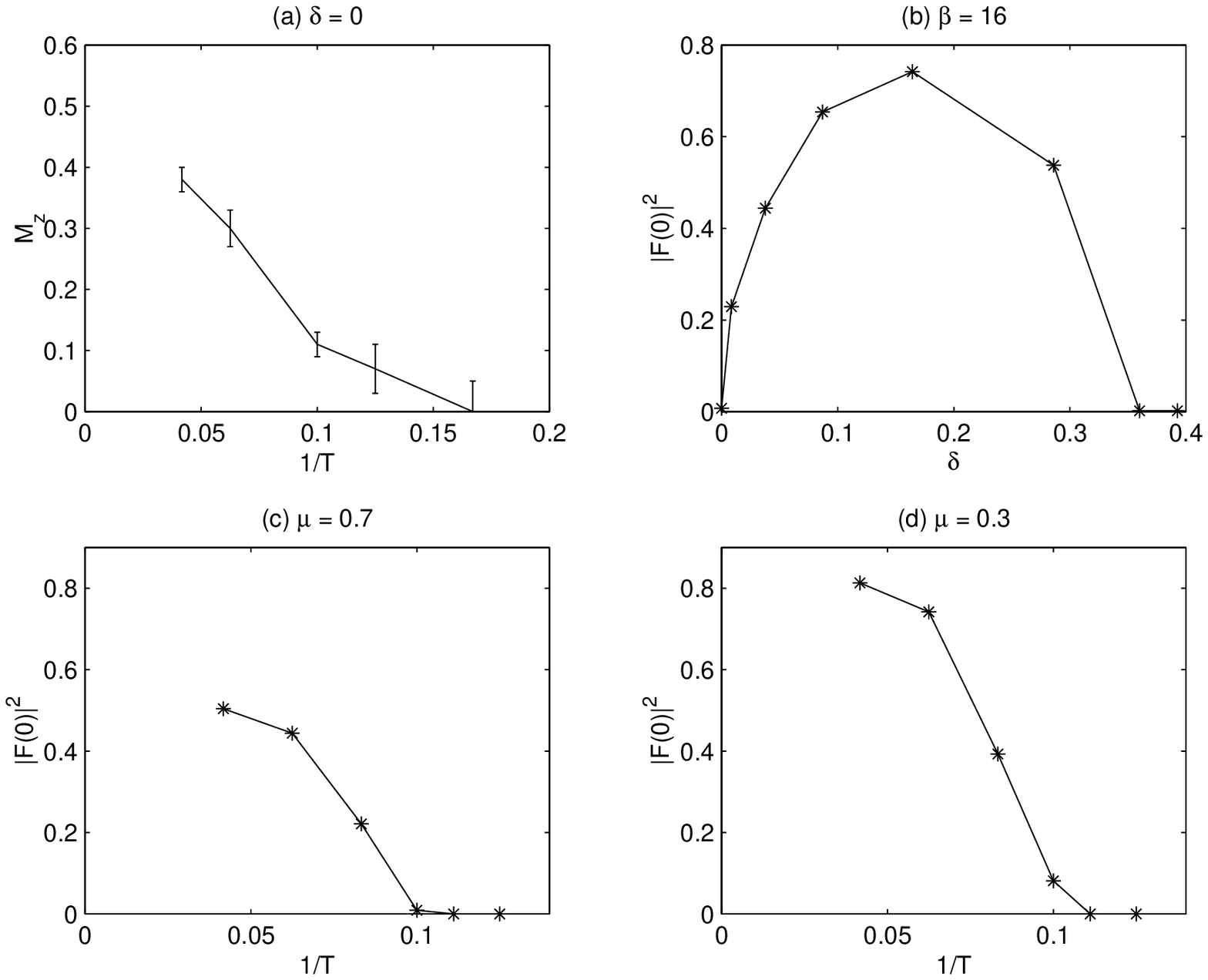}
{\bf Figure 13 :} (a)$M_{z}$ vs. temperature at half filling. Units are
$\mu_{B}$ and {2t} for vertical and horizontal axis, respectively.
(b)$|F(0)|^{2}$ vs. doping at $\beta = 16$. A unit for the vertical axis
is arbitrary. $|F(0)|^{2}$ vs. temperature for $\mu = 0.7$(c)  and $\mu =
0.3$(d). 
\end{figure}

\begin{figure}
\includegraphics{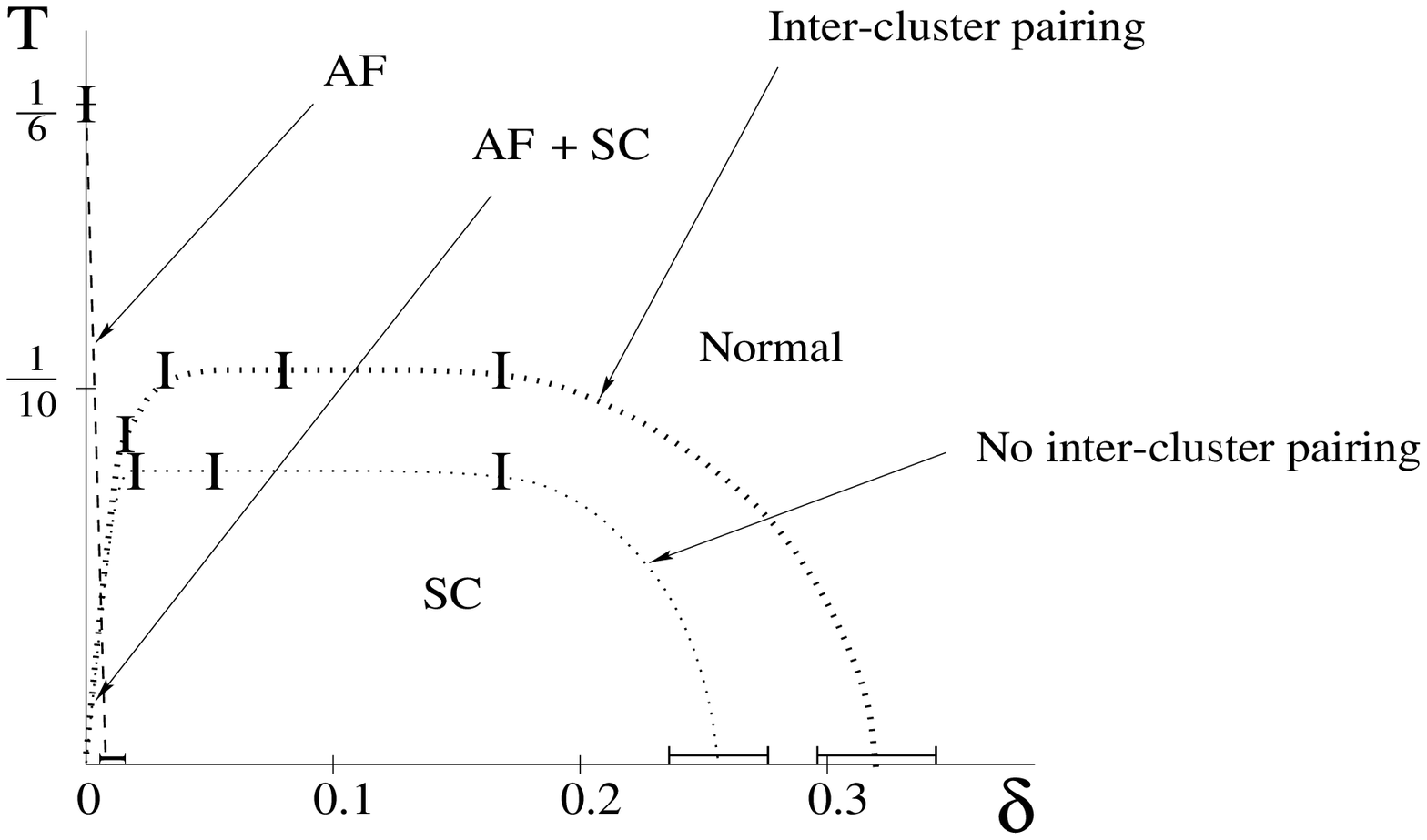}
{\bf Figure 14 :}Schematic phase diagram obtained based on our model.
The onset of SC was determined by the appearance of a
thermal average of an instantaneous pairing amplitude. Error bars
correspond to the size of increments of sampling points we studied. 
\end{figure}


\begin{thebibliography}{54}
\bibitem{watanabe4} A metal-insulator transition in the presence of \AFc order where carrier doping is not involved has been analyzed based on the two-band Hubbard model in DMFT\cite{watanabe1, watanabe2}.
\bibitem{watanabe6} We found that for $\mu=0.7$, when a narrow peak at $\omega = 0$ in the pair hopping channel
(the weight of this peak basically corresponds to the Meissner weight) starts to develop
(below $\beta = 10$), the pair breaking channel is roughly gapped $\sim
\Delta_{1\:particle} \ge \frac{k_{B}T}{t}$ ($= \frac{1}{\beta t} \sim
0.2$) and thus suppressed. On the other hand, for $\mu=0.3$ when weight
starts to develop at $\omega = 0$ in the pair-hopping channel, the pair
breaking channel still has significant weight in the low energy sector
$\le \frac{k_{B}T}{t}$, which implies that the amplitude fluctuations are
significantly enhanced for $\mu=0.3$ near $T_{c}$. Since the energy scale
for pair breaking exactly matches to the single-particle gap(Fig.4), the
distinction between $\mu=0.7$ and $\mu=0.3$ essentially comes from a
difference in the robustness of a single-particle gap\cite{watanabe2}.
\end{thebibliography}
\end{document}